\documentclass[useAMS,usenatbib]{mn2e}
\usepackage{natbib}
\usepackage{amssymb,amsmath,amsfonts}
\usepackage{epsfig}
\usepackage{mathptmx}
\usepackage{longtable}
\usepackage{graphicx}
\usepackage{comment}
\usepackage{color}
\usepackage[caption=false]{subfig}
\usepackage{animate}
\usepackage[T1]{fontenc}
\usepackage{aecompl}

\def\reff@jnl#1{{\rm#1\/}}

\def\aj{\reff@jnl{AJ}}                  
\def\araa{\reff@jnl{ARA\&A}}            
\def\apj{\reff@jnl{ApJ}}                        
\def\apjl{\reff@jnl{ApJ}}               
\def\apjs{\reff@jnl{ApJS}}              
\def\ao{\reff@jnl{Appl.Optics}}         
\def\apss{\reff@jnl{Ap\&SS}}            
\def\aap{\reff@jnl{A\&A}}               
\def\aapr{\reff@jnl{A\&A~Rev.}}         
\def\aaps{\reff@jnl{A\&AS}}             
\def\azh{\reff@jnl{AZh}}                        
\def\baas{\reff@jnl{BAAS}}              
\def\jrasc{\reff@jnl{JRASC}}            
\def\memras{\reff@jnl{MmRAS}}           
\def\mnras{\reff@jnl{MNRAS}}            
\def\pra{\reff@jnl{Phys. Rev. A}}         
\def\prb{\reff@jnl{Phys. Rev. B}}         
\def\prc{\reff@jnl{Phys. Rev. C}}         
\def\prd{\reff@jnl{Phys. Rev. D}}         
\def\prl{\reff@jnl{Phys. Rev. Lett}}      
\def\pasp{\reff@jnl{PASP}}              
\def\pasj{\reff@jnl{PASJ}}              
\def\qjras{\reff@jnl{QJRAS}}            
\def\skytel{\reff@jnl{S\&T}}            
\def\solphys{\reff@jnl{Solar~Phys.}}    
\def\sovast{\reff@jnl{Soviet~Ast.}}     
\def\ssr{\reff@jnl{Space~Sci.Rev.}}     
\def\zap{\reff@jnl{ZAp}}                        
\def\nat{\reff@jnl{Nature}}             

\def\p#1by#2{{\partial{#1} \over \partial{#2}}}
\def\pp#1by#2#3{{\partial^2{#1} \over \partial{#2}\partial{#3}}}
\def\d#1by#2{{{\rm d}{#1} \over {\rm d}{#2}}}
\def\dd#1by#2#3{{{\rm d}^2{#1} \over {\rm d}{#2}{\rm d}{#3}}}

\title[]{Deep radio observations of the radio halo of the bullet cluster 1E\,0657-55.8} 

\label{firstpage}

\author[Shimwell et~al.]{Timothy W. Shimwell$^1$\thanks{E-mail: Timothy.Shimwell@csiro.au},
 Shea Brown$^2$,
 Ilana J. Feain$^1$,
 Luigina Feretti$^3$,
 \newauthor
 B. M. Gaensler$^{4}$, 
 Craig Lage$^5$\\ \\
  $^1$ CSIRO Astronomy \& Space Science, Australia Telescope National Facility, PO Box 76, Epping, NSW 1710, Australia \\
 $^2$ Department of Physics and Astronomy, University of Iowa, 203 Van Allen Hall, Iowa City, IA 52242, U.S.A \\
 $^3$ INAF - Istituto di Radioastronomia, via Gobetti 101, 40129 Bologna, Italy \\
$^4$ Sydney Institute for Astronomy, School of Physics, The University of Sydney, NSW 2006, Australia \\
 $^5$ Center for Cosmology and Particle Physics, Department of Physics, New York University, NY, NY 10003, USA\\
\date{Accepted ---; received ---; in original form \today}}
\begin{document}
\maketitle
\begin{abstract}

We present deep 1.1-3.1\,GHz Australia Telescope Compact Array observations of the radio halo of the bullet cluster, 1E\,0657-55.8. In comparison to existing images of this radio halo the detection in our images is at higher significance.  The radio halo is as extended as the X-ray emission in the direction of cluster merger but is significantly less extended than the X-ray emission in the perpendicular direction. At low significance we detect a faint second peak in the radio halo close to the X-ray centroid of the smaller sub-cluster (the bullet) suggesting that, similarly to the X-ray emission, the radio halo may consist of two components. Finally, we find that the distinctive shape of the western edge of the radio halo traces out the X-ray detected bow shock. The radio halo morphology and the lack of strong point-to-point correlations between radio, X-ray and weak-lensing properties suggests that the radio halo is still being formed. The colocation of the X-ray shock with a distinctive radio brightness edge illustrates that the shock is influencing the structure of the radio halo. These observations support the theory that shocks and turbulence influence the formation and evolution of radio halo synchrotron emission.

\end{abstract}


\section{Introduction}

Low surface brightness, steep spectrum, diffuse radio emission has been observed to be associated with the intra-cluster medium (ICM) of many merging galaxy cluster systems (see \citealt{Bruggen_2012}, \citealt{Feretti_2012}  and \citealt{Ferrari_2008} for recent reviews). The emission is due to synchrotron radiation from relativistic electrons gyrating in a magnetic field. When coincident with the centre of the merging cluster system, this emission is referred to as a radio halo, whereas if located on the periphery of the cluster it is known as a radio relic -- \cite{Kempner_2004} provides details on the classification of extended radio sources in galaxy clusters. While radio halos are believed to be the result of cluster-wide processes such as post-merger turbulence (see e.g. \citealt{Brunetti_2001} and \citealt{Petrosian_2001}) or proton-proton collisions (see e.g. \citealt{Dennison_1980}), radio relics are believed to be associated in some way with localised, post-merger shock-fronts (\citealt{Ensslin_1998}). Measurements of X-ray shocks in the regions of radio relics (see e.g. \citealt{Macario_2011}, \citealt{Bourdin_2013}, \citealt{Giacintucci_2008}, \citealt{Ogrean_2013a},  \citealt{Akamatsu_2013} and \citealt{Ogrean_2013b}) and synchrotron radio spectral index gradients (see e.g. \citealt{Weeren_2010}) have provided strong support to the theory that these objects are formed by first-order Fermi acceleration (diffusive shock acceleration). The exact connection/relationship between relics and halos is unknown, but some observations of radio halos have shown that weak X-ray shocks are approximately cospatial with edges of radio halos (see e.g. \citealt{Brown_2011} and \citealt{Markevitch_2012}). An example of such a system is 1E 0657-55.8 (usually referred to as `the bullet cluster' because of its distinctive morphology). In this case the radio images of the powerful radio halo indicate that it has an irregular shape (\citealt{Liang_2000}) and that the position of the radio edge and the X-ray shock is similar, however, it was unclear from the halo observations by \cite{Liang_2000} whether there was any correspondence between the morphology of the distinctive X-ray bow shock and the (indistinct) radio halo edge. We have obtained observations that are deeper than those of \cite{Liang_2000} and are able to characterise the halo in greater detail and at higher resolution. The primary aim of this study is to determine whether shocks influence the structure of radio halos.

The bullet cluster with its prominent bow shock and bright radio halo is an obvious source for investigating the relationship between shocks and radio halos. We have used the Australia Telescope Compact Array (ATCA) to obtain deep, large fractional bandwidth (1.1-3.1\,GHz), polarimetric observations of the bullet cluster which we use to compare the radio emission from relativistic electrons with the X-ray emission from thermal gas. 

Hereafter we assume a concordance $\rm{\Lambda}$CDM cosmology, with $\rm{\Omega_{m}}$ = 0.3, $\rm{\Omega_\Lambda}$ = 0.7 and h $\equiv$ H$_{0}$/(100 km\,s$^{-1}$Mpc$^{-1}$) = 0.7. At $z=0.296$, the luminosity distance of 1E\,0657-55.8 is 1529 Mpc and 1$\arcsec$ corresponds to 4.413\,kpc. All coordinates are given in J2000.

\section{Observations}

\subsection{Archival radio data}\label{Sec:previous-atca}

\cite{Liang_2000} presented a serendipitous discovery of a radio halo associated with 1E 0657-55.8 and using 128\,MHz bandwidth ATCA observations centred at 1.3, 2.4, 4.9, 5.9 and 8.8\, GHz as well as 843\,MHz Molonglo Observatory Synthesis Telescope (MOST) observations they characterised its properties. By integrating over a 3.5\,Mpc$^2$ region of the cluster, they measured that the halo has a total flux density of 78$\pm$5\,mJy at 1.3\,GHz. In a central 0.78\,Mpc$^2$ region, where the halo is brightest, they measured an integrated flux density of $\approx$50\,mJy at 1.3\,GHz and a 0.8-8.8\,GHz spectral index $\alpha=-1.3\pm$0.1 (where $I_{\nu} \propto {\nu}^\alpha$) with no significant spectral steepening between these frequency limits. Combining the large area integrated flux measurement and smaller scale spectral index measurement they calculated the rest frame 1.4\,GHz power of the halo to be 4.3$\pm$0.3$\times10^{25}$\,W\,Hz$^{-1}$. The cosmology used by \cite{Liang_2000} assumed that 1$\arcsec$ corresponds to 5.45\,kpc, with this cosmology 0.78 and 3.5\,Mpc$^2$ correspond to 7.3 and 32.7 square arcminutes respectively.

\cite{Liang_2000} did not detect polarisation in the radio halo but derived upper limits of 20\%, 6.5\% and 1.4\% at resolutions of 10$\arcsec$, 20$\arcsec$ and 60$\arcsec$ from 1.3\,GHz images. The authors noted the similarity between the X-ray and radio halo surface brightness distributions but added that there appeared to be a tighter correspondence between the halo emission and the optical galaxy distribution.

The thermal noise in the \cite{Liang_2000} images at a resolution of 60$\arcsec$ is 1100, 51, 110, 56, 65 and 56\,$\mu$Jy/beam at frequencies of 0.8, 1.3, 2.4, 4.9, 5.9 and 8.8\,GHz respectively. The 1.3\,GHz data were also mapped with resolutions of 6.5$\arcsec$ and 23$\arcsec$; the noise levels on these images were 44\,$\mu$Jy/beam and 90\,$\mu$Jy/beam respectively.

\subsection{Recent radio data}

Since the observations presented by \cite{Liang_2000} were performed, the ATCA has undergone a significant upgrade in which the bandwidth, spectral resolution and system temperature have all been improved (see \citealt{Wilson_2011}).

\begin{figure}
   \centering
   \includegraphics[width=8cm]{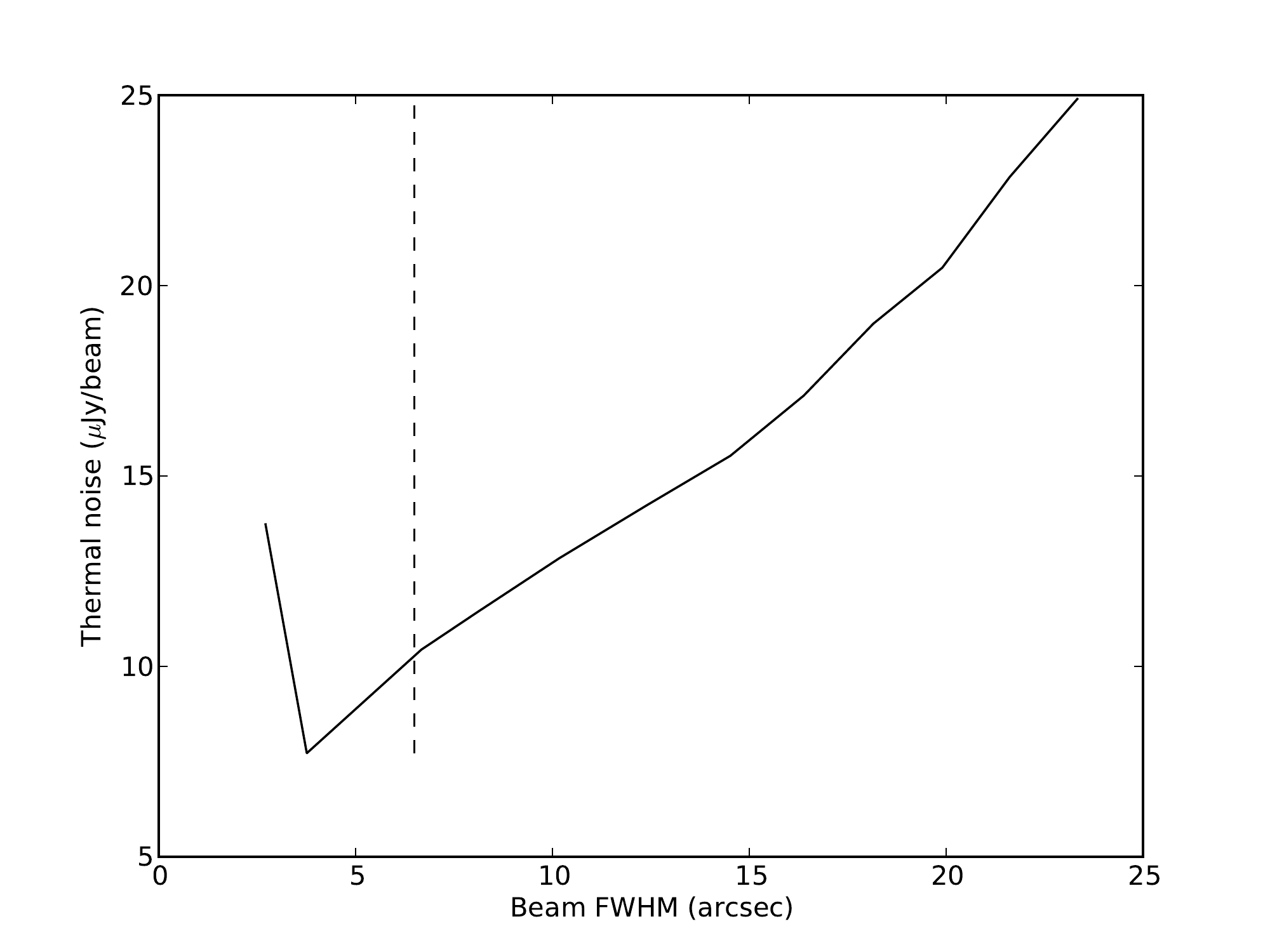} 
   \caption{The measured thermal noise of the 1.1-3.1\,GHz ATCA observations towards 1E 0657-55.8 as a function of angular resolution. The natural resolution (marked with a dashed vertical line) of our observations is 6.5$\arcsec$ and at this resolution the sensitivity is 10$\mu$Jy/beam.}
   \label{fig-angular-sensitivity}
\end{figure}

We used the ATCA to observe 1E 0657-55.8 on 2012 December 17 and 22, 2013 January 2, 2013 February 17 ,and 2014 January 8 and 10 (project C2756). ATCA observations of our target were centred at the coordinates used by \cite{Liang_2000} and were interleaved every 30 minutes with 5 minute phase-calibrator observations. For amplitude calibration, PKS B1934-638 was observed for 10 minutes in each of the observing sessions. We observed in the 1.5B,1.5D, 6A and 6B configurations to provide a compromise between high surface brightness sensitivity and angular resolution. Long-duration observations in each configuration gave good $uv$-coverage and sensitivity to different angular scales -- the measured thermal noise as a function of resolution for our observations of the bullet cluster is shown in Figure \ref{fig-angular-sensitivity}. The noise continues to decrease a little beyond the diffraction limit as the sidelobes are suppressed. The shortest baseline with which we have observed is 31\,m (corresponding to angular scales of 30.2$\arcmin$ at 1.1\,GHz or 10.7$\arcmin$ at 3.1\,GHz) and the longest baseline used was 5970\,m (corresponding to 9.4$\arcsec$ at 1.1\,GHz and 3.3$\arcsec$ at 3.1\,GHz). A summary of the observations is provided in Table \ref{ATCA-obs}.

\begin{table}
\caption{A summary of our ATCA observations towards 1E 0657-55.8. The quoted
  synthesised beam FWHM and sensitivity correspond to a natural weighting of the
  visibilities, however the data were imaged at a variety of resolutions between 2.7$\arcsec$ and $23.3\arcsec$. Figure \ref{fig-angular-sensitivity} demonstrates how the sensitivity varies as a
  function of synthesised beam size for these data. The observations were carried out in four configurations of the ATCA antennas; the target was observed for 8, 5, 9 and 10 hours in the 1.5B, 1.5D, 6A and  6B arrays, respectively. The measured polarisations allow for Stokes I, Q, U and V imaging.}
 \centering
 \label{ATCA-obs}
\begin{tabular}{lccc}
\hline 
Coordinates (J2000) & 06:58:32.7 -55:57:19.0 \\
Amplitude calibrator & PKS B1934-638 \\
Phase calibrator & PMN J0742-56 \\
Receiver noise & 14$\mu$Jy/beam\\
On-source time & 32 hours \\
Frequency range & 1.1-3.1\,GHz\\
Spectral resolution & 1\,MHz \\
Synthesised beam FWHM & 6.5$\arcsec$ natural resolution \\
Primary beam FWHM & 42$\arcmin$-15$\arcmin$ \\
Polarisations measured & XX, YY, XY and YX \\ \hline
 \end{tabular}
\end{table}

\section{Data reduction}

\subsection{Calibration}\label{sec:calibration}

The data from the target, amplitude calibrator and phase calibrator were excised of data from bad channels and radio frequency interference (RFI) using the \textsc{miriad} implementation of AOFlagger (\citealt{Offringa_2010}) and additional manual flagging. The remaining data were then calibrated using \textsc{miriad}\footnote{http://www.atnf.csiro.au/computing/software/miriad/} (\citealt{Sault_1995}). The stable calibrator source, PKS B1934-638, which has no linear or circular polarisation and a flux-density as a function of frequency as characterised by \cite{Reynolds_1994}, was used to perform a calibration of the complex gains across the bandpass and to determine the polarisation leakages. The phase calibrator, PMN J0742-56, was used to refine calibration parameters and to update those that vary with time.

The Stokes I data were further calibrated in \textsc{Miriad} with four iterations of phase self-calibration followed by a single iteration of phase-plus-amplitude self-calibration. This procedure requires an accurate \textsc{clean} component model, which was created by multi-frequency \textsc{clean}ing the Stokes I data with visibilites weighted such that the synthesised beam size was 4$\arcsec$. The \textsc{clean} component model was built up slowly through the multiple imaging iterations. Initially the model contained only the most significant \textsc{clean} components, but after the full number of iterations \textsc{clean} components from regions in which the flux-density exceeded 5 and 3 times the thermal noise were used for the phase self calibration and phase-plus-amplitude self-calibration, respectively.

\subsection{Continuum imaging}\label{sec:continuum_imaging}

After self-calibration in \textsc{miriad}, the data were imaged in \textsc{CASA}\footnote{http://casa.nrao.edu/} using the multi-scale multi-frequency deconvolution algorithm (\citealt{Rau_2011}) to accurately map the complex and extended structure of the 1E 0657-55.8 radio halo and surrounding region. The images were \textsc{clean}ed to five times the thermal noise with deconvolution scale sizes equal to zero (point source), one and three times the synthesised beam. This procedure was performed to create full bandwidth (1.1-3.1\,GHz) Stokes I images.

\begin{figure}
   \centering
   \includegraphics[width=8cm]{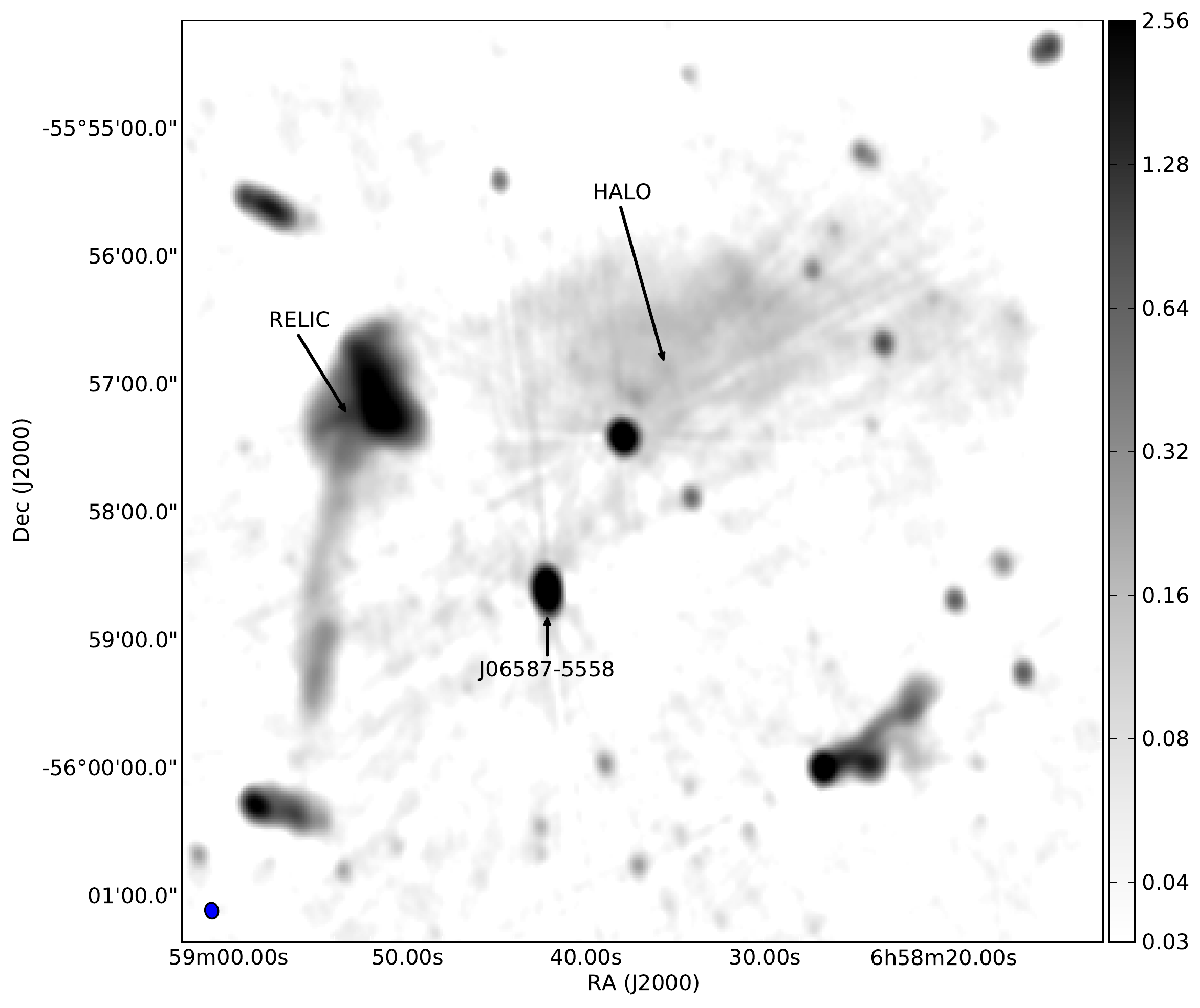} 
   \caption{A primary beam corrected, medium resolution (synthesised beam FWHM $\approx$ 7$\arcsec$), Stokes I, 1.1-3.1\,GHz image towards 1E 0657-55.8. The greyscale is in mJy/beam and the measured thermal noise is 11$\mu$Jy/beam. The radio halo, LEH2001\,J06587-5558 and the proposed radio relic are labelled.}
   \label{greyscale-image}
\end{figure}

The 1E 0657-55.8 radio halo has a linear angular extent of $>4\arcmin$ and the data were imaged with various visibility weightings to examine how its observed structure changes as the synthesised beam FWHM is varied from 2.7$\arcsec$ to $23.3\arcsec$. The robust parameter within the \cite{Briggs_1995} weighting scheme was used to apply a more uniform weighting to visibilities across the $uv$-plane, and a Gaussian taper was applied to reduce the weight of the data from the long baselines. Figure \ref{fig-angular-sensitivity} shows the measured sensitivity as a function of synthesised beam area and Figure \ref{greyscale-image} shows a medium resolution 1.1-3.1\,GHz image of the dataset.

\subsection{Radio source detection and subtraction}\label{sec:subtraction}

To study the faint diffuse emission from the radio halo, it is vital that contaminating sources are identified and removed from the data. To isolate such sources from the radio halo, we imaged the data from baselines longer than 4000$\lambda$ at high resolution (FWHM=2.7$\arcsec$) to resolve out the extended radio halo and used the \textsc{sourcefind} software package (\citealt{Franzen_2011}) to detect sources with flux-densities greater than  5$\sigma$ in this image, where $\sigma$ varies across the map due to confusion noise and noisy regions around bright sources  (see \citealt{Shimwell_2013}).

\begin{figure}
   \centering
   \includegraphics[width=8cm]{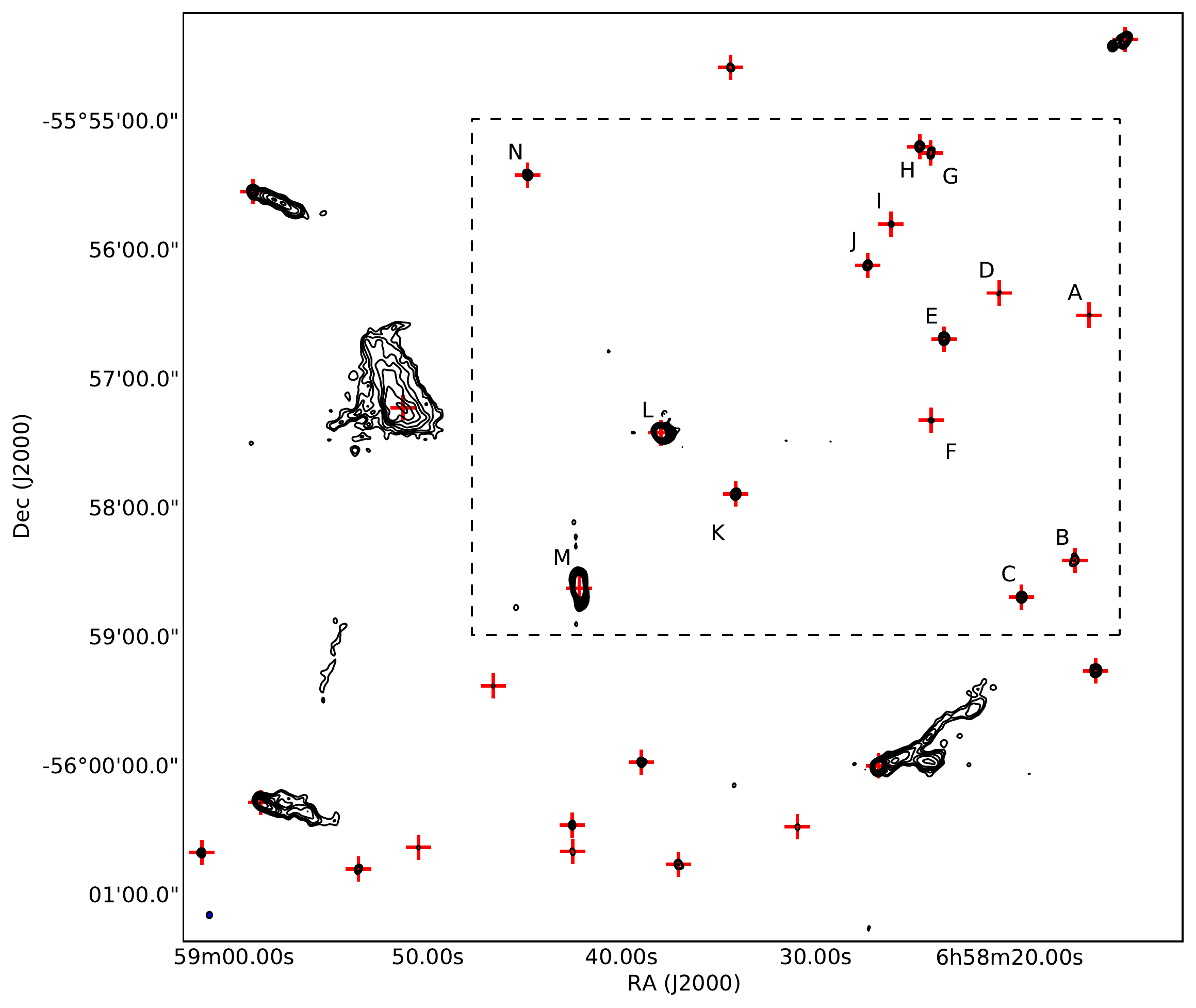} 
   \caption{The primary beam corrected Stokes I image with the highest resolution  (synthesised beam FWHM $\approx$ 2.7$\arcsec$) at which we have mapped the data. Data from baselines shorter than 4000$\lambda$ (corresponding to 50$\arcsec$) have been excluded to resolve out the extended halo. The dashed rectangle shows the \textsc{clean} box in which we subtract the \textsc{clean} components that were created in this high resolution imaging process. This image is before the \textsc{clean} components have been subtracted. The contour levels are at $\sqrt{1,2,4,8,...}\times 5 \times 15\mu$Jy/beam. The locations of detected sources are shown with crosses and the properties of sources whose emission contaminates the halo are given in Table \ref{tab:ATCA_SOURCES}.}
   \label{fig:casa-imagedays_0_1_9_14all-bandrobust-2-deeper_cleanbox}
\end{figure}

We identified a total of 57 sources within 6$\arcmin$ (at $z$=0.296 this corresponds to 1.59\,Mpc) of the pointing centre and of these 43 are extended, where we define extended as sources whose area exceeds 1.4 $\times$ the area of the synthesised beam. Fourteen sources lie within the halo (see Table \ref{tab:ATCA_SOURCES} and Figure \ref{fig:casa-imagedays_0_1_9_14all-bandrobust-2-deeper_cleanbox}) and the \textsc{casa} tasks \textsc{ft} and \textsc{uvsub} were used to subtract the \textsc{clean} components of these contaminating sources from the visibilities. Unfortunately residuals remain on the images after subtraction (see Table \ref{tab:ATCA_SOURCES}), and although these residuals are small, they contaminate our images of the halo. Therefore, on all source subtracted images we mark the positions of subtracted sources, so that the reader can treat these areas with caution.

\begin{table*} 
\caption{Parameters of the 14 radio sources with flux densities exceeding 5$\sigma$, which contaminate the 1.1-3.1\,GHz emission from the radio halo. The wide-band flux measurements were obtained from the 2.7$\arcsec$ resolution primary beam corrected 1.1-3.1\,GHz image shown in Figure \ref{fig:casa-imagedays_0_1_9_14all-bandrobust-2-deeper_cleanbox}. The sources have been subtracted from the wide-band data following the procedure described in Section \ref{sec:subtraction} and the residual contamination after subtraction is given in this Table. Images of the 1.1-3.1\,GHz data before \textsc{clean} component subtraction are shown in Figures \ref{greyscale-image} and \ref{fig:casa-imagedays_0_1_9_14all-bandrobust-2-deeper_cleanbox} and images after \textsc{clean} component subtraction are shown in Figures \ref{fig-bullet-cluster} and \ref{fig-bullet-cluster-resolution}. For sources detectable in 290\,MHz sub-bands (see Section Section \ref{sec:trad-spec}) we have derived spectral index parameters and 2.1\,GHz flux-density measurements from narrow-band primary beam corrected images that have been convolved to all have an equal resolution of 6.9$\arcsec$. }
 \label{tab:ATCA_SOURCES} 
\begin{tabular}{lcccccccccccc} 
\hline 

Source & Right  & Declination & 1.1-3.1\,GHz  & 1.1-3.1\,GHz  & Extended &  Residual   & $\alpha^{3.1}_{1.1}$ & 2.1\,GHz & 2.1\,GHz \\  
ID &   Ascension &                & Peak & Integrated       &        at 2.7$\arcsec$                                      & 1.1-3.1\,GHz &  & Peak & Integrated \\ 
                     &      &            & Flux-density & Flux-density &  Resolution                                    & Flux-density &  & Flux-density & Flux-density \\ 
                  &       &          &     (mJy) & (mJy)              &               &                             ($\mu$Jy)  & & (mJy) & (mJy)\\ \hline
A & 06:58:15.9 & -55:56:30.6 & $0.09\pm0.02$ & $0.15\pm0.02$ & $\surd$ & 70 & -- & -- & -- \\ 
B & 06:58:16.6 & -55:58:24.7 & $0.16\pm0.02$ & $0.46\pm0.03$ & $\surd$ & 110 & -- & -- & -- \\ 
C & 06:58:19.3 & -55:58:41.7 & $0.63\pm0.02$ & $0.67\pm0.02$ & $\times$ & 90 & $-0.5\pm0.1$ & $0.6\pm0.1$ & $0.6\pm0.1$ \\ 
D & 06:58:20.5 & -55:56:20.3 & $0.08\pm0.02$ & $0.19\pm0.02$ & $\surd$ & 90 & -- & -- & -- \\ 
E & 06:58:23.4 & -55:56:41.7 & $0.78\pm0.02$ & $0.88\pm0.02$ & $\times$ & 100 & $-1.00\pm0.08$ & $0.7\pm0.1$ & $0.8\pm0.1$ \\ 
F & 06:58:24.0 & -55:57:19.6 & $0.12\pm0.02$ & $0.13\pm0.02$ & $\times$ & 50 & -- & -- & -- \\ 
G & 06:58:24.0 & -55:55:15.2 & $0.21\pm0.02$ & $0.35\pm0.02$ & $\surd$ & 90 & -- & -- & -- \\ 
H & 06:58:24.6 & -55:55:12.2 & $0.38\pm0.02$ & $0.46\pm0.02$ & $\times$ & 90 & $-1.3\pm0.1$ & $0.4\pm0.1$ & $0.4\pm0.1$ \\ 
I & 06:58:26.1 & -55:55:48.3 & $0.12\pm0.02$ & $0.14\pm0.02$ & $\times$ & 90 & -- & -- & -- \\ 
J & 06:58:27.3 & -55:56:07.4 & $0.27\pm0.02$ & $0.37\pm0.02$ & $\surd$ & 80 & $-0.6\pm0.2$ & $0.3\pm0.1$ & $0.3\pm0.2$ \\ 
K & 06:58:34.1 & -55:57:53.7 & $0.50\pm0.02$ & $0.59\pm0.02$ & $\times$ & 100 & $-0.8\pm0.1$ & $0.5\pm0.2$ & $0.5\pm0.2$ \\ 
L & 06:58:37.9 & -55:57:25.4 & $8.18\pm0.02$ & $14.66\pm0.03$ & $\surd$ & 130 & $-0.82\pm0.01$ & $9.4\pm0.2$ & $13.7\pm0.2$ \\ 
M & 06:58:42.1 & -55:58:37.6 & $7.74\pm0.02$ & $25.75\pm0.04$ & $\surd$ & 100 & $-0.93\pm0.01$ & $9.8\pm0.2$ & $24.0\pm0.2$ \\ 
N & 06:58:44.8 & -55:55:25.4 & $0.47\pm0.02$ & $0.52\pm0.02$ & $\times$ & 50 & $0.6\pm0.1$ & $0.5\pm0.1$ & $0.5\pm0.2$ \\                   \hline
\end{tabular} 
\end{table*}

 \subsection{Spectral index measurements}\label{sec:spectral_index}

We derived spectral index measurements using the following two methods. \\
\\
1)  Wide-band method: multi-scale multi-frequency \textsc{clean} was used to image the entire 1.1-3.1\,GHz band using two Taylor terms to describe the sky brightness as a function of frequency \\ \\
2)  Narrow-band method: the 1.1-3.1\,GHz band was split into 290\,MHz sub-bands, and multi-scale multi-frequency \textsc{clean}ing was applied to each sub-band image separately. \\ \\
The first approach makes better use of the data; due to the more complete $uv$-coverage, deconvolution using all 1.1-3.1\,GHz channels will be more reliable than that obtained in any single 290\,MHz sub-band. However, images with such a large fractional bandwidth are not frequent in the literature and thus we also use the narrow-band method, a more traditional spectral index imaging technique, to reassure readers that the wide-band method is valid for these data. We created spectral index images with different resolutions by weighting the visibilities appropriately (see Section \ref{sec:continuum_imaging}).

\subsubsection{Wide-band method}\label{sec:rau_method}

The multi-frequency imaging that we described in Section \ref{sec:continuum_imaging} was performed with two Taylor polynomial terms to describe the sky brightness as a function of frequency,
\begin{equation}
I_{\nu} =  I_{0} + I_{1} \left( \frac{\nu-\nu_{0}}{\nu_{0}} \right),
\label{eqn:taylor-exp-spectral}
\end{equation}
where $I_{0}$ and $I_{1}$ are images that correspond to the first two coefficients in the Taylor expansion of the sky brightness, $\nu$ is the frequency under consideration and $\nu_0$ is the reference frequency which we set to 2.1\,GHz. By assuming that sources are described with a spectral index, $\alpha$, such that 
\begin{equation}
I_{\nu} = I_{\nu_{0}} \left( \frac{\nu}{\nu_0} \right)^\alpha,
\label{eqn:spectral-index}
\end{equation}
a comparison of Equation \ref{eqn:taylor-exp-spectral} with the Taylor expansion around  $I_{\nu_{0}}$  of Equation \ref{eqn:spectral-index} gives  $I_{\nu_0} =  I_{0}$  and $\alpha=\frac{I_{1}}{I_{\nu_0}}$. Images of both  $I_{\nu_0}$ and $\alpha$ were produced during the multi-frequency multi-scale imaging (see \citealt{Rau_2011} for a thorough description of this algorithm). This algorithm also produces an $\alpha_{error}$ image, where the errors are dependent upon the bandwidth, number of independent measurements, the $uv$-coverage and the signal-to-noise ratio. To obtain accurate spectra away from the pointing centre, we used the task \textsc{widebandpbcor}, which uses a polynomial fit to the ATCA primary beam to make corrections to flux estimates prior to the creation of spectral index images. We note that CASA is capable of using additional Taylor terms to model the spectral curvature; for example, if three Taylor terms are used, the spectrum of a source is described as $I_{\nu} = I_{\nu_{0}} \left( \frac{\nu}{\nu_0} \right)^{\alpha + \beta log(\nu/\nu_{0})}$. However, using additional Taylor terms requires high signal to noise. Little spectral curvature is expected for 1E\,0657-55.8 in any case, since \cite{Liang_2000} showed that, within the error bars, there is no spectral curvature between 0.843\,GHz and 8.8\,GHz for this radio halo.

\subsubsection{Narrow-band method}\label{sec:trad-spec}

We split the 1.1-3.1\,GHz band into successive 290\,MHz sub-bands, and imaged each of these sub-bands with the same pixel size and a $uv$-tapering that corresponds to a desired resolution  before correcting for the primary beam. This produces a cube containing images at different frequencies but with approximately the same resolution; thus for each pixel we have flux measurements in each 290\,MHz sub-band. Every plane of the cube was convolved with a Gaussian to give all planes exactly the same resolution. To create a spectral index map we found, for each pixel, the power-law that best described the flux density of that pixel as a function of frequency. We propagated the errors in our flux estimates to errors in the calculated spectral index, and then blanked out pixels with errors larger than 0.3. Additionally, we blanked pixels where the pixel flux did not exceed five times the noise in all 290\,MHz sub-bands.

\subsection{Polarisation imaging and rotation measure synthesis}

We have used two techniques to characterise the polarisation properties. In the first method we use rotation measure synthesis to search for polarised emission; since maximum sensitivity is desired for this, we use the entire wide-band 1.1-3.1\,GHz dataset. In the second method we split the 1.1-3.1\,GHz band into successive 100\,MHz sub-bands and image each sub-band separately to determine the polarised intensity as a function of frequency for polarised objects. We use the pre-source subtracted dataset (see Section \ref{sec:subtraction}) for both of these polarisation analyses.

\subsubsection{Rotation measure synthesis}\label{sec:rm-cube}

The procedure we have used for rotation measure synthesis  (\citealt{Brentjens_2005}) is the same as was used by \cite{OSullivan_2012}. A thorough description is presented in that paper but here we present a summary of the procedure.

After calibration of the 1.1-3.1\,GHz dataset (see Section \ref{sec:calibration}) the Stokes I, Q, U and V data were imaged in 10\,MHz sub-bands, in this process each sub-band was imaged with approximately the same resolution and \textsc{clean}ed to five times the thermal noise. The images were then each convolved with a Gaussian to give them exactly the same resolution. The resulting Q and U image cubes were used to calculate the Faraday dispersion function, which describes the complex polarised surface brightness per unit Faraday depth and is calculated for each pixel ($\alpha,\delta$):
\begin{equation}
F(\phi,\alpha,\delta) \approx \frac{\sum_{i=1}^{N} w_i  P_{i,\alpha,\delta} e^{-2i \phi (\lambda_{i}^{2} -\lambda_{0}^2)}}{\sum_{i=1}^{N}w_i},
\end{equation}
where the sum is over channels, $\phi$ is the Faraday depth, $\alpha$ is the right ascension, $\delta$ is the declination, $w_i P_{i,\alpha,\delta}$ is the weighted complex polarised intensity (P = Q + iU ) for a pixel and $\lambda_{0}$ is the weighted mean $\lambda^2$ over all channels -- we find $\lambda_{0}$=0.15m. The weighting is 
$w_i = \frac{1}{\sigma_{Q,U}^2}$, where $\sigma_{Q,U}$ is the thermal noise level of the appropriate Q and U image. 

\begin{figure}
   \centering
   \includegraphics[width=8cm]{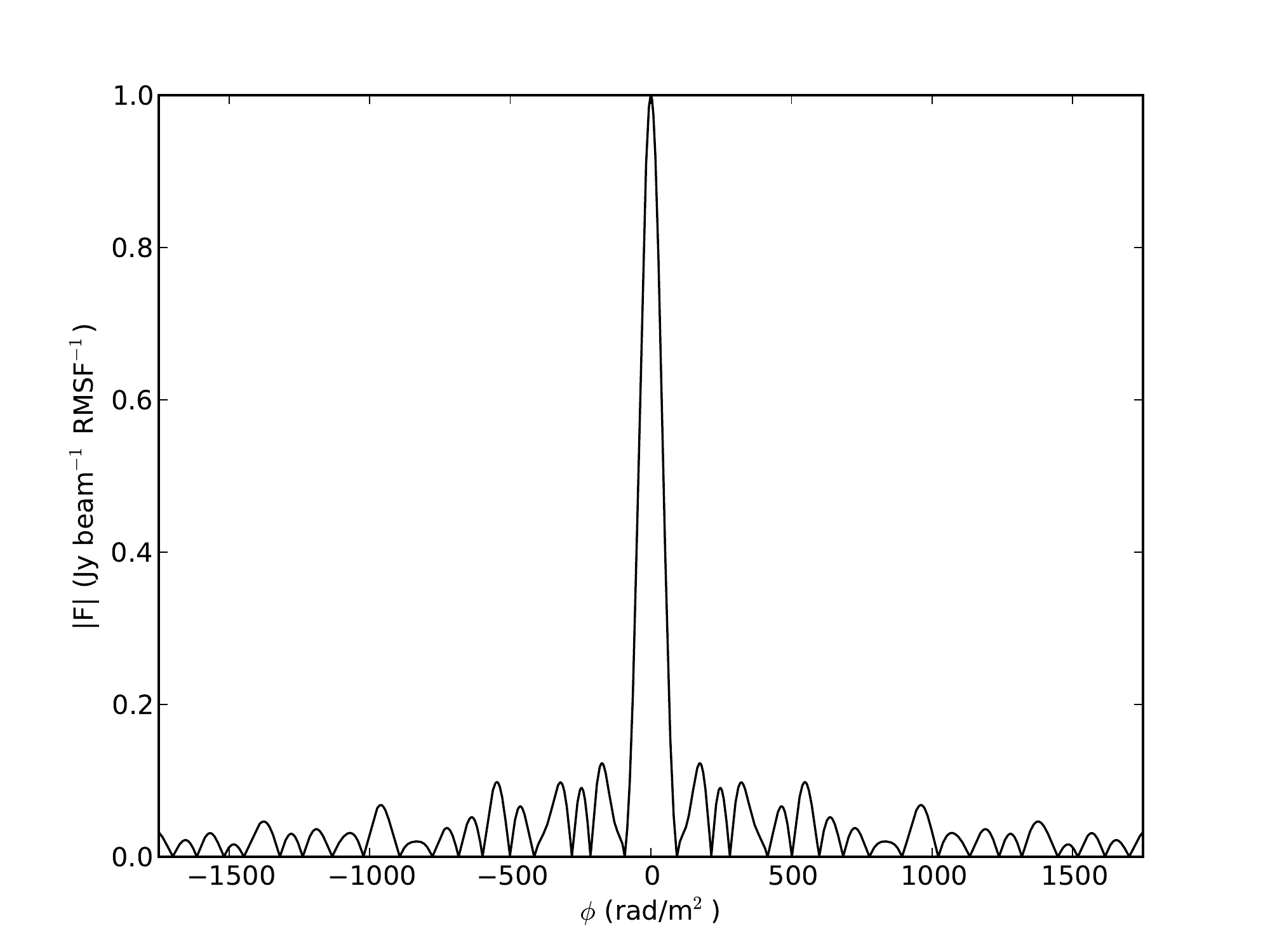} 
   \caption{The amplitude of the rotation measure spread function for the 1.1-3.1\,GHz ATCA polarisation data.}
   \label{fig:rmsf}
\end{figure}

The Faraday dispersion function was calculated for integer values of $\phi$ between -1750 and 1750\,rad/m$^2$ to produce an Faraday depth cube, with each slice showing the complex polarisation for a different value of $\phi$. Figure \ref{fig:rmsf} shows the amplitude of the rotation measure spread function (RMSF) for our 1.1-3.1\,GHz dataset, in which bad data have been flagged and 10\,MHz sub-bands have been weighted according to their thermal noise levels. For our analysis of the data, the FWHM of the RMSF is 84\,rad/m$^{2}$, the largest detectable scale is 320\,rad/m$^{2}$ and the maximum observable Faraday depth is 1650\,rad/m$^{2}$.

 As we simply want to detect polarisation, we have not attempted to use \textsc{RMCLEAN} (\citealt{Heald_2009}). Additionally, we do not take into account the spectral index of the Stokes I intensity in our calculation of the Faraday dispersion function (see Section 3 of \citealt{Brentjens_2005}).

\subsubsection{Polarisation as a function of frequency}\label{sec:pol_freq}

To study polarisation variations of faint sources as a function of frequency, we must compromise between the bandwidth required to make sensitive measurements and the effects of bandwidth depolarisation. We split the wide-band 1.1-3.1\,GHz dataset into multiple 100\,MHz sub-bands, and for each of these sub-bands we created images of the Stokes Q and U data. Polarised intensity images were created from these images ($P= \sqrt{Q^2 + U^2}$). The polarisation images were corrected for the Ricean bias and pixels with polarised intensities of less than 2$\sigma_{Q,U}$ were flagged. To determine the polarisation as a function of frequency for a source, we integrated the pixels within the region of the source that had a polarised flux greater than five times the noise (where we use $\sigma_{Q,U}$ for the noise). This procedure was performed on each of the sub-band polarisation images to measure polarisation changes as a function of frequency. The errors that we provide on this measurement are $\sigma_{Q,U}$ multiplied by the square root of the source area in synthesised beams.

\section{Results}

In Figure \ref{greyscale-image} we show the 1.1-3.1\,GHz image of the bullet cluster, 1E 0657-55.8. Faint large-scale emission is observed just north of the centre of the image; this is the radio halo and is the primary focus of this paper. To the south-east of the radio halo is LEH2001\,J06587-5558 (labelled as J06587-5558), this source was studied in  \cite{Liang_2001} but we have used our wide-band observations to provide some new measurements of its properties. The bright source with a large north-south extension that lies on the eastern edge of the radio halo has the spectral and polarimetric properties expected for a radio relic, but we leave a detailed discussion of this source to a forthcoming publication (Shimwell et al., in preparation). Three tailed radio sources are visible to the south, south-east and north-east of the radio halo (see Figure \ref{greyscale-image} and the higher resolution image shown in Figure \ref{fig:casa-imagedays_0_1_9_14all-bandrobust-2-deeper_cleanbox}), and a further tailed radio source is located just outside the image at 06:58:43.4 -56:02:35.3 -- these tailed radio sources are not discussed further in this publication.

\subsection{Radio halo}

\begin{figure*}
\centering
\includegraphics[width=17cm]{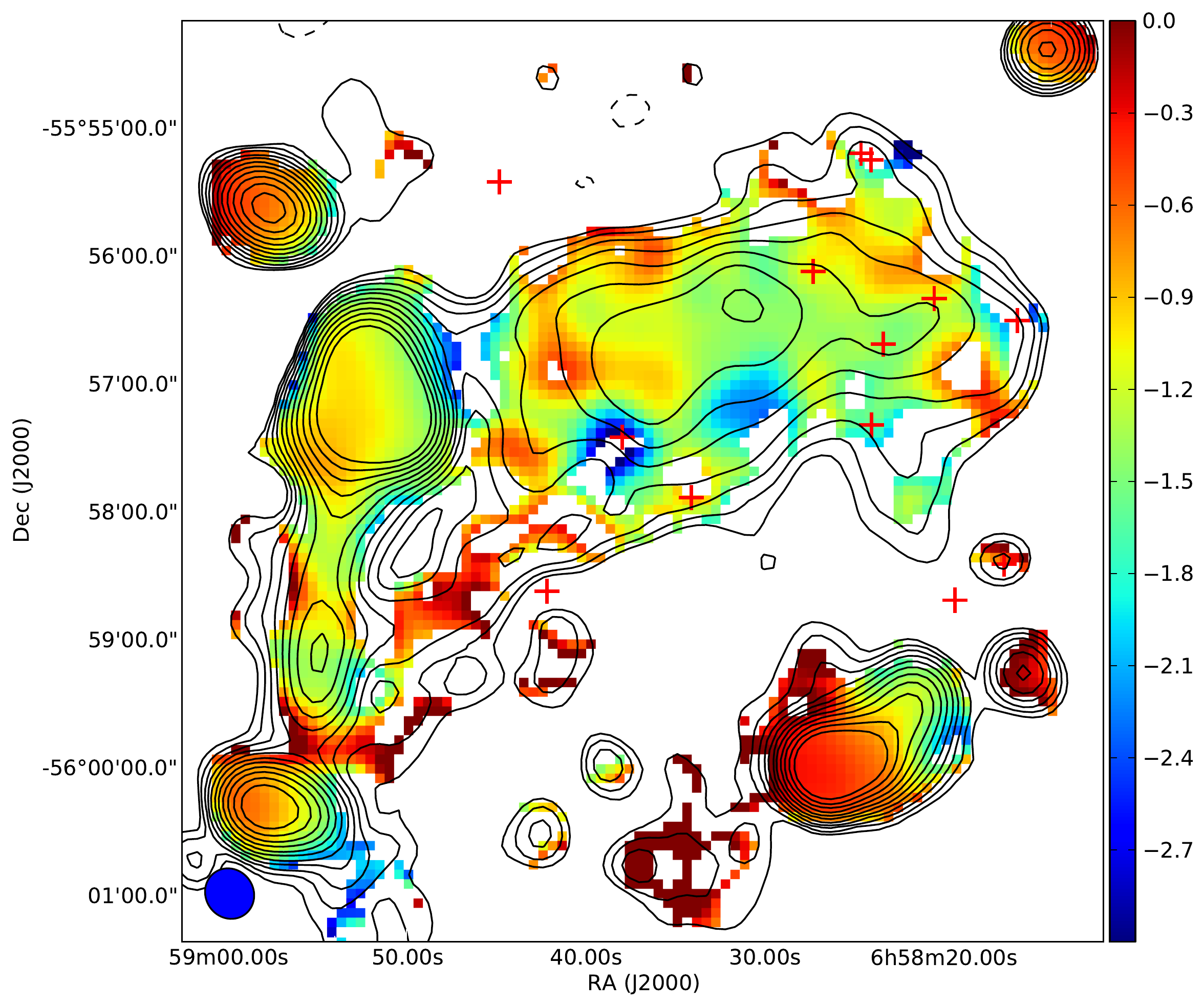}
\includegraphics[height=5.2cm]{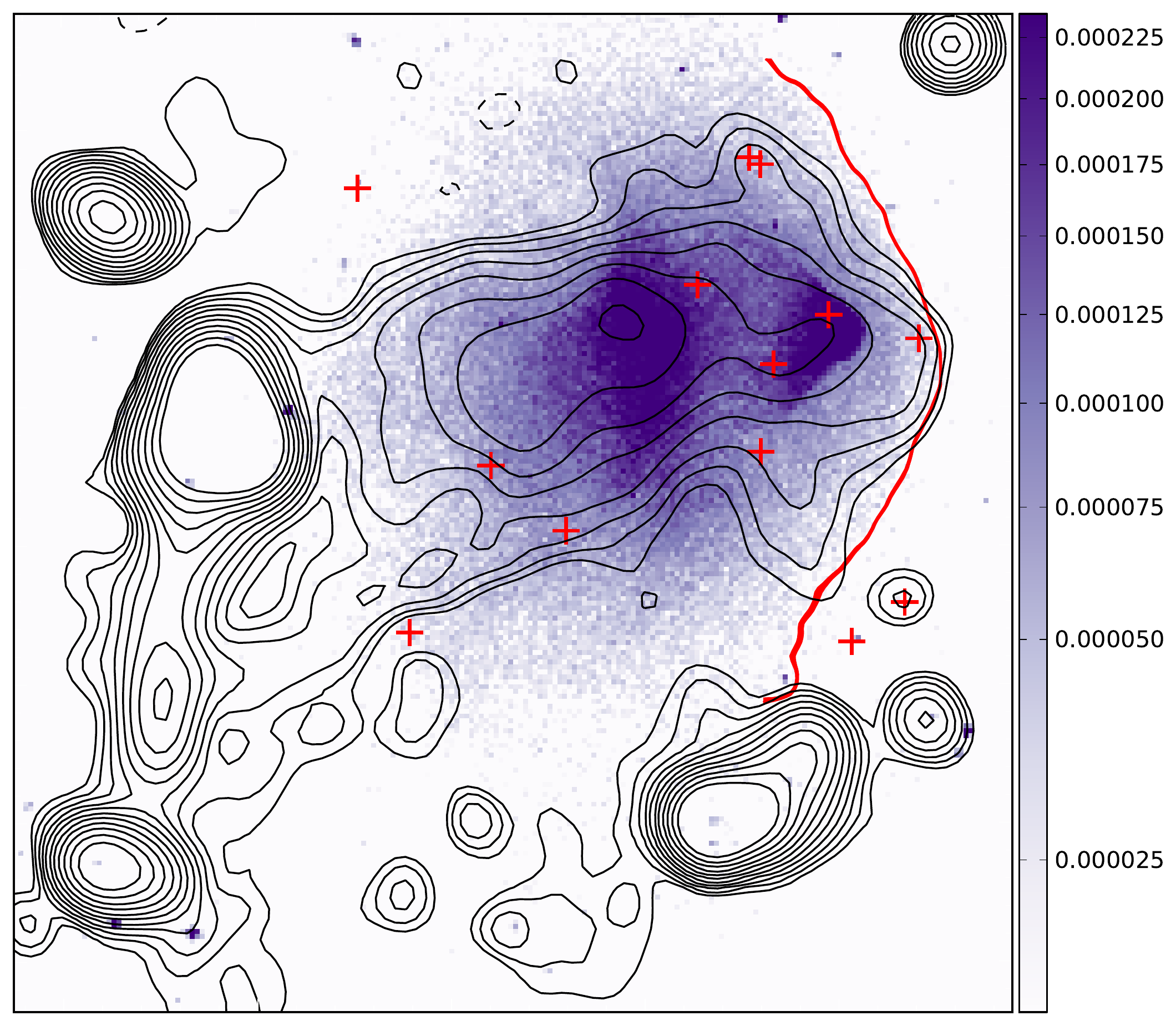}
\includegraphics[height=5.2cm]{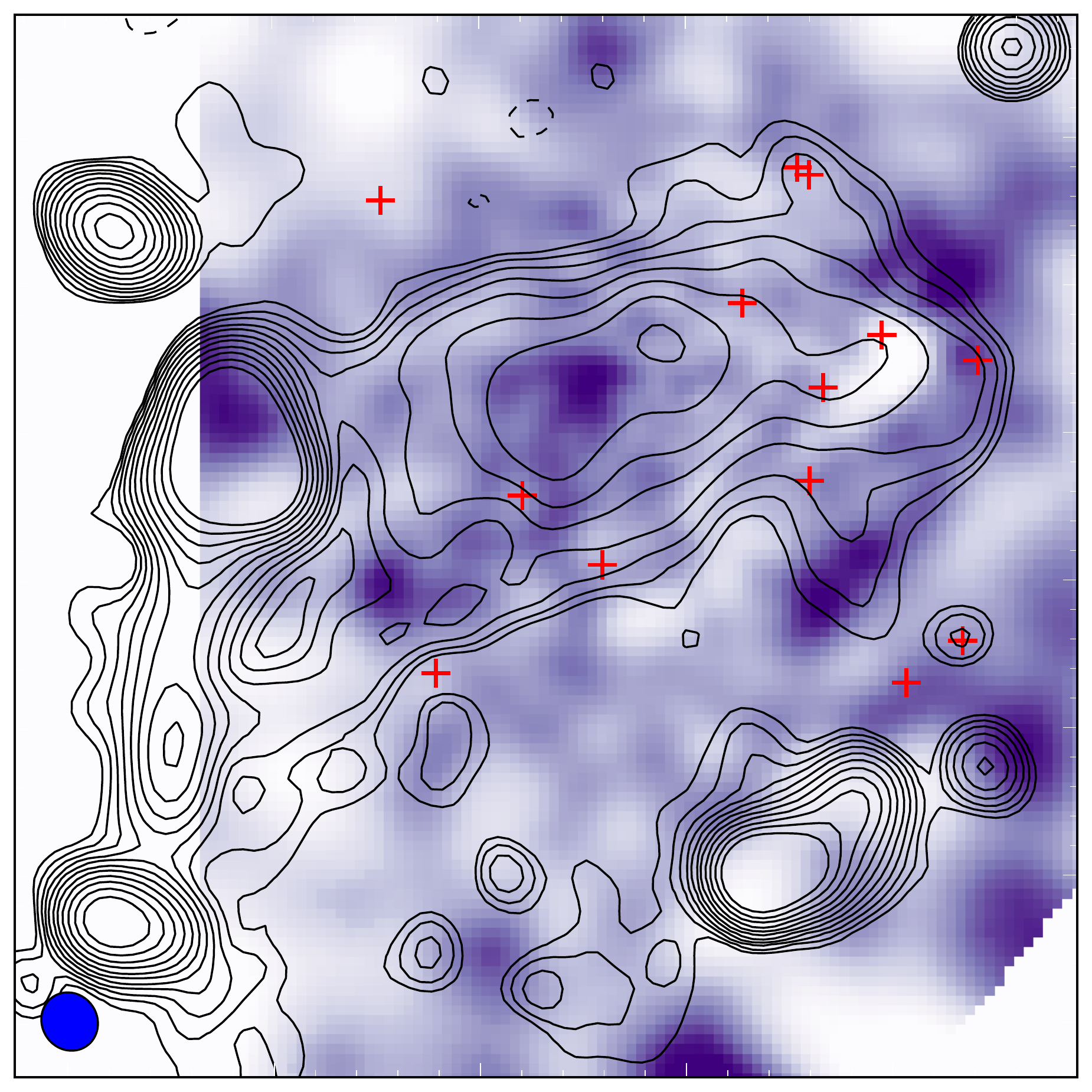}
 \includegraphics[height=5.2cm]{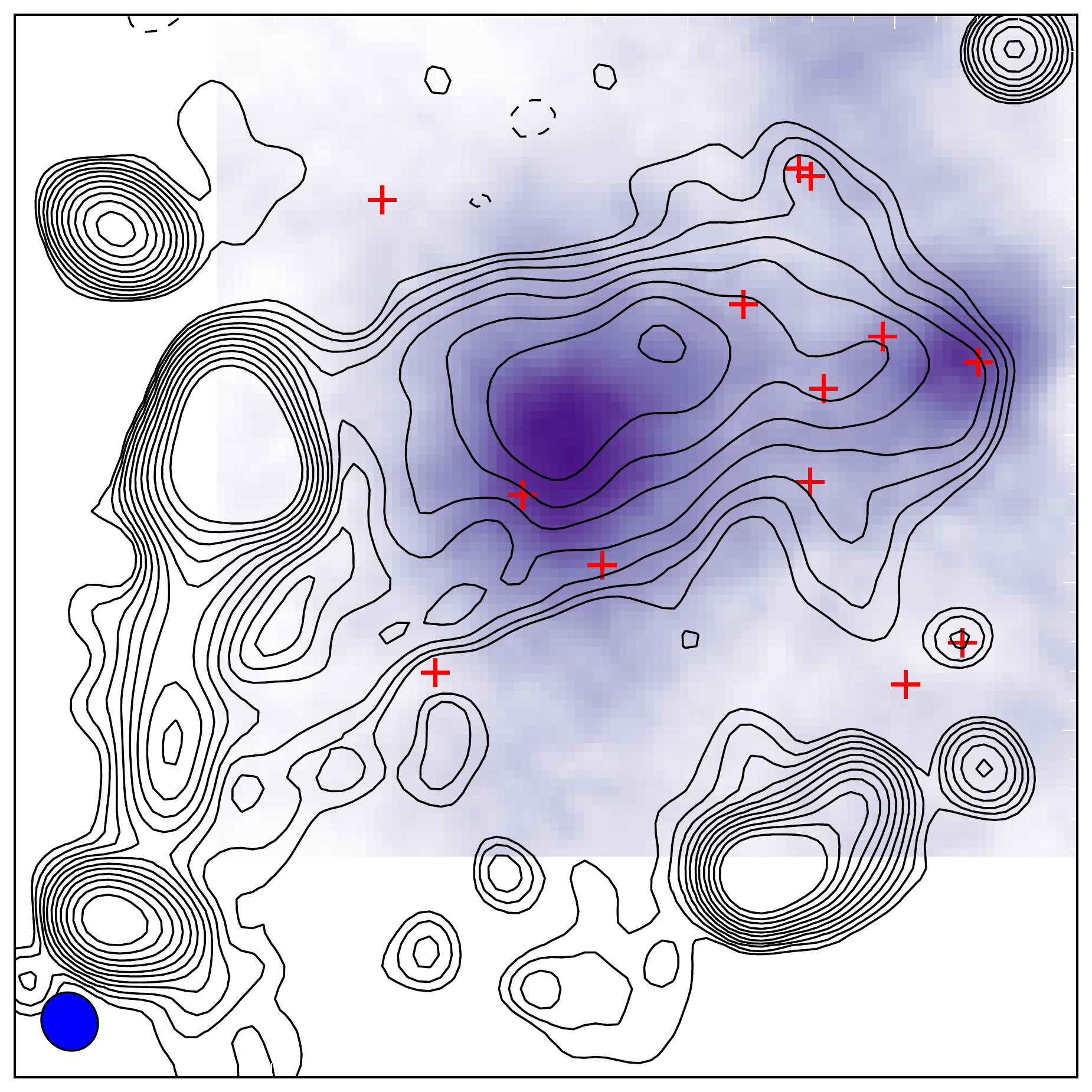} 
   \caption{Top: the contours show the 1.1-3.1\,GHz primary beam corrected continuum emission of the radio halo at a resolution of $23.3\arcsec$. The colour scale shows the 1.1-3.1\,GHz spectral index image at the same resolution which was created using the wide-band method described in Section \ref{sec:rau_method}. All regions on the spectral index image where the error exceeds 0.3 have been blanked and the errors on this 1.1-3.1\,GHz spectral index image are presented in Figure \ref{fig-bullet-cluster-error}. Images at different resolutions are shown in Figure \ref{fig-bullet-cluster-resolution}. Bottom left: low resolution ATCA contours overlaid on a colour scale image made from 500\,ks of \textsc{Chandra} ACIS-I data from the 0.8-4\,keV band (see \citealt{Markevitch_2006}). The scale is in cts s$^{-1}$pixel$^{-1}$ (the pixel size is 1.968$\arcsec$) and the red line shows the X-ray shock front (as defined by \citealt{Chung_2009}, i.e. tracing out constant X-ray intensity from 06:58:24.0 -55:54:06.0 to 06:58:24.0 -55:44:28.8). Bottom centre: low resolution ATCA data overlaid on an X-ray temperature image with the scale in keV (Markevitch 2014, in preparation; see also \citealt{Owers_2009}). Bottom right: low resolution ATCA contours overlaid on a colour scale corresponding to the weak lensing reconstruction presented by \protect\cite{Clowe_2006}. The colour scale is in weak lensing $\kappa$ units and can be converted to solar masses per pixel (the pixel size is 3.552$\arcsec$) with a conversion factor of 7$\times 10^{11}$M$_{\odot}$pixel$^{-1}$. The weak lensing data are from http://flamingos.astro.ufl.edu/1e0657/public.html. For all images the ellipse in the bottom left corner shows the ATCA synthesised beam and the contour levels are 5 $\times \sqrt{1,2,4,8,...}\times 25\mu$Jy/beam; positive contours are solid lines and negative contours are dashed. All images have been primary beam corrected.}
 \label{fig-bullet-cluster}
\end{figure*}

\begin{figure*}
\centering
\includegraphics[height=5.2cm]{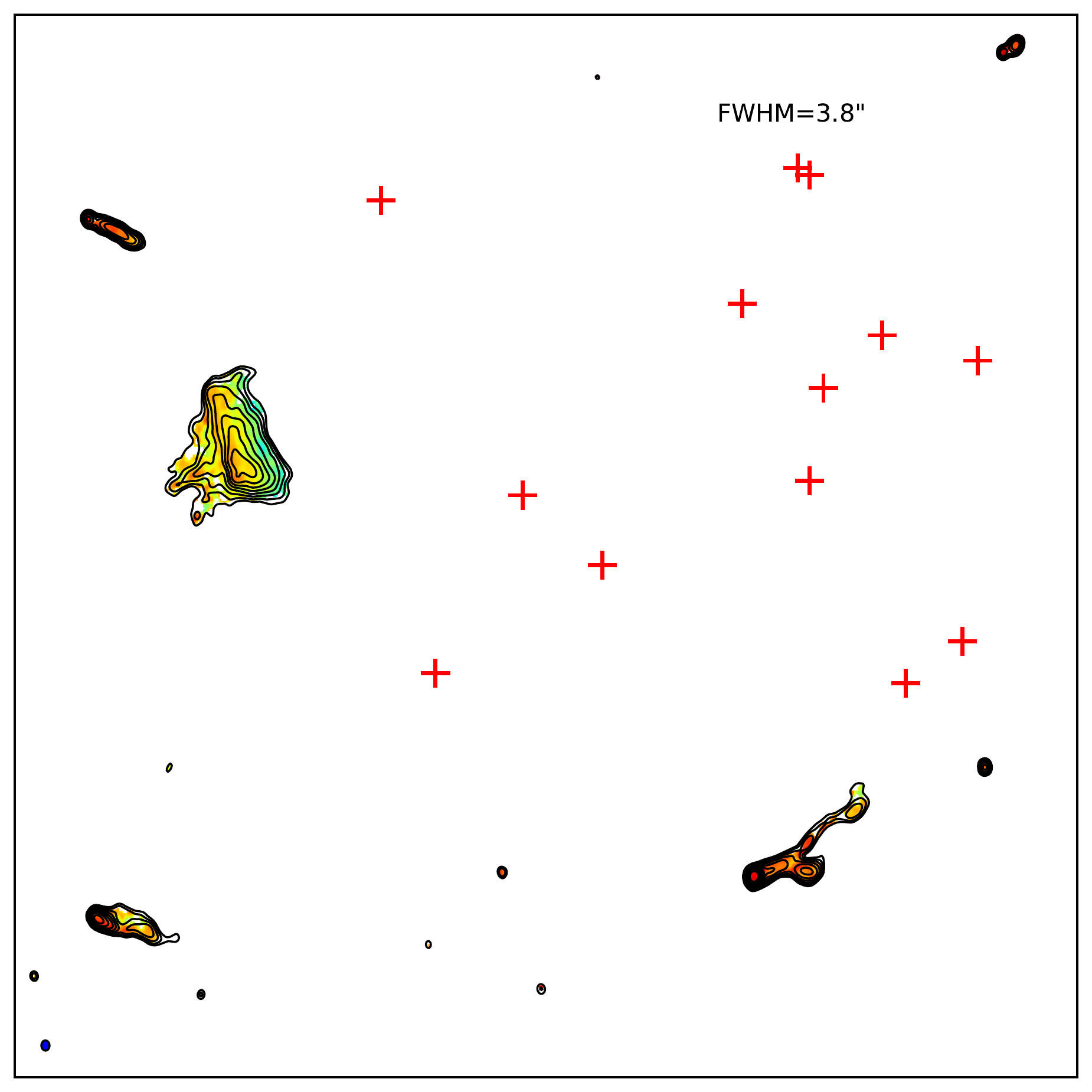}
\includegraphics[height=5.2cm]{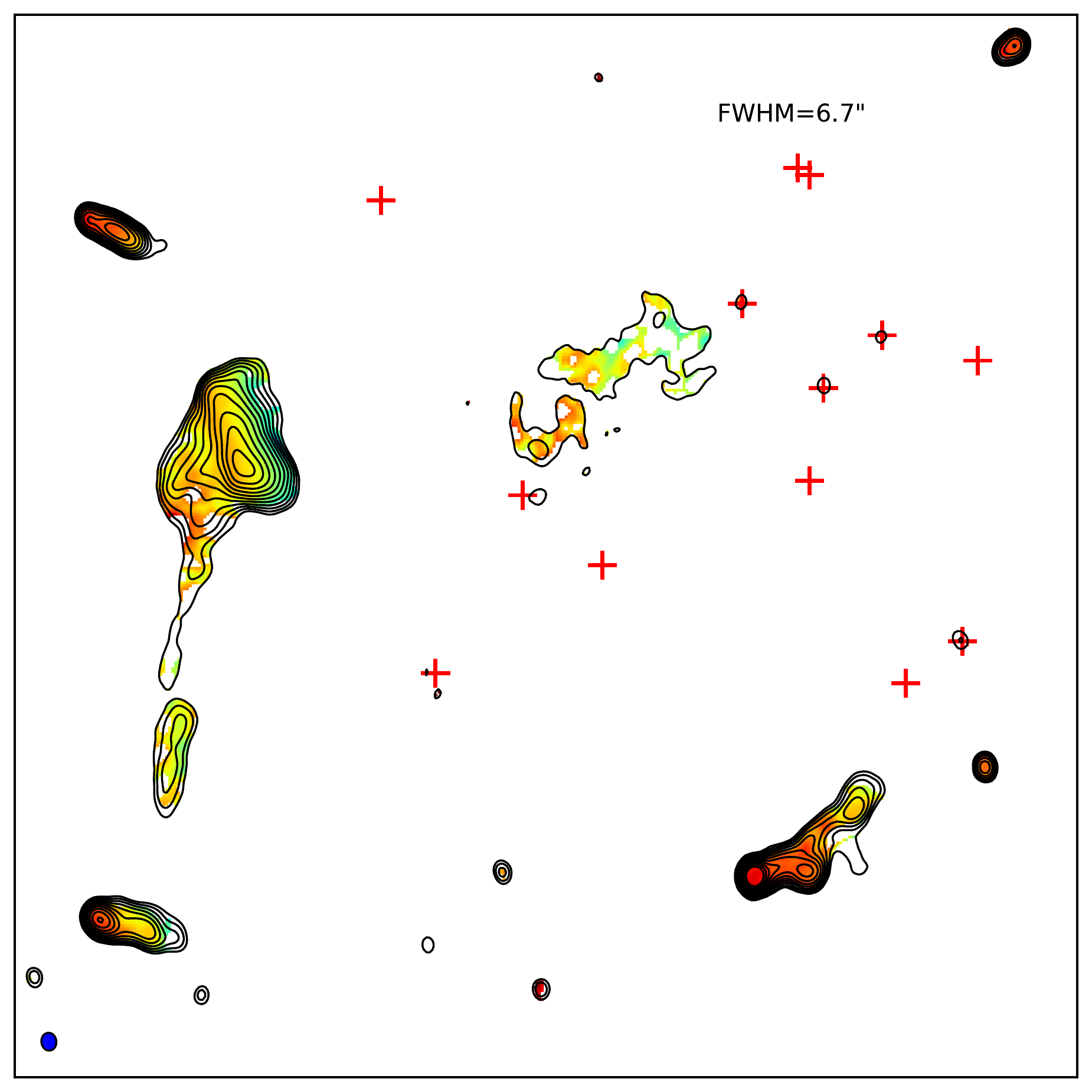}
\includegraphics[height=5.2cm]{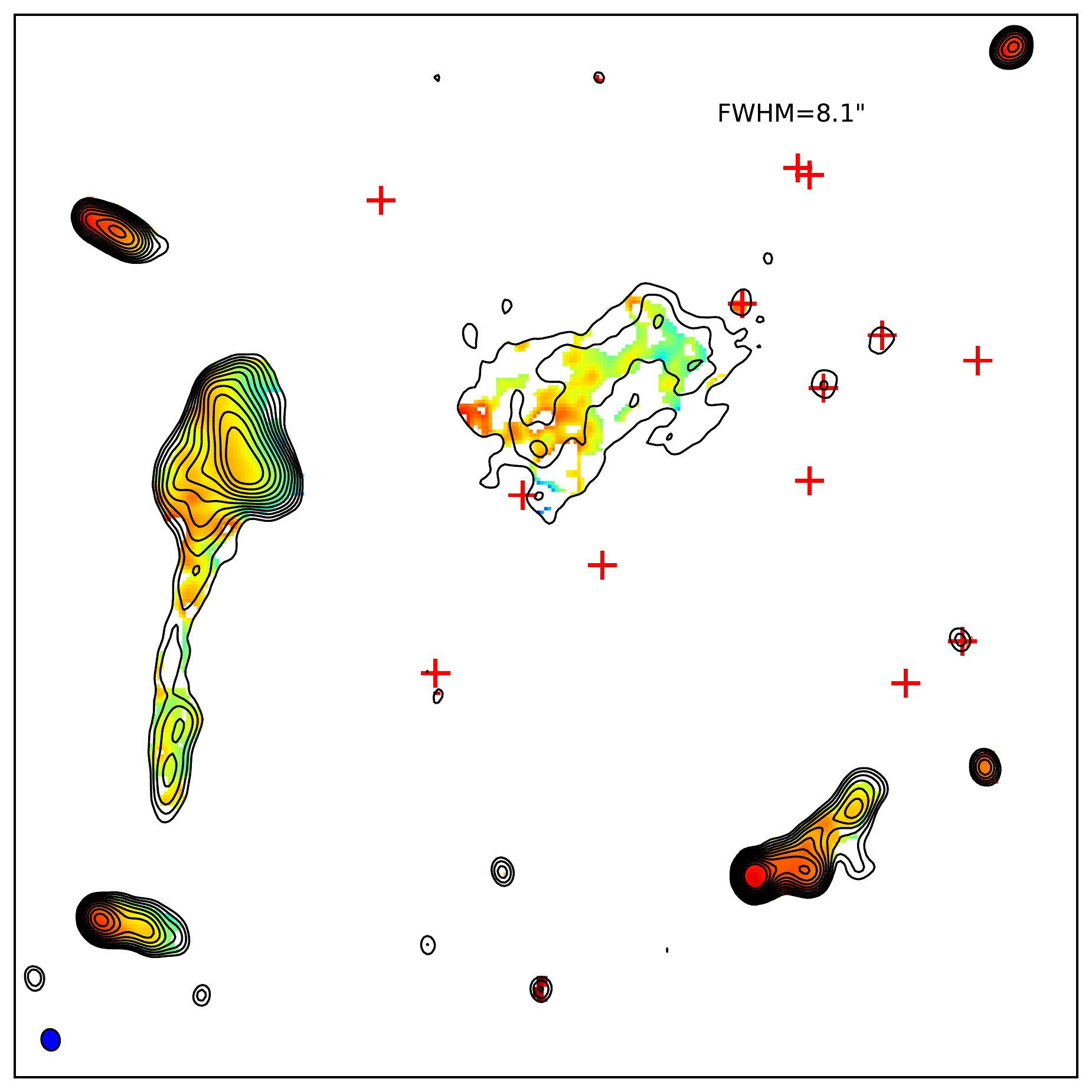}
\includegraphics[height=5.2cm]{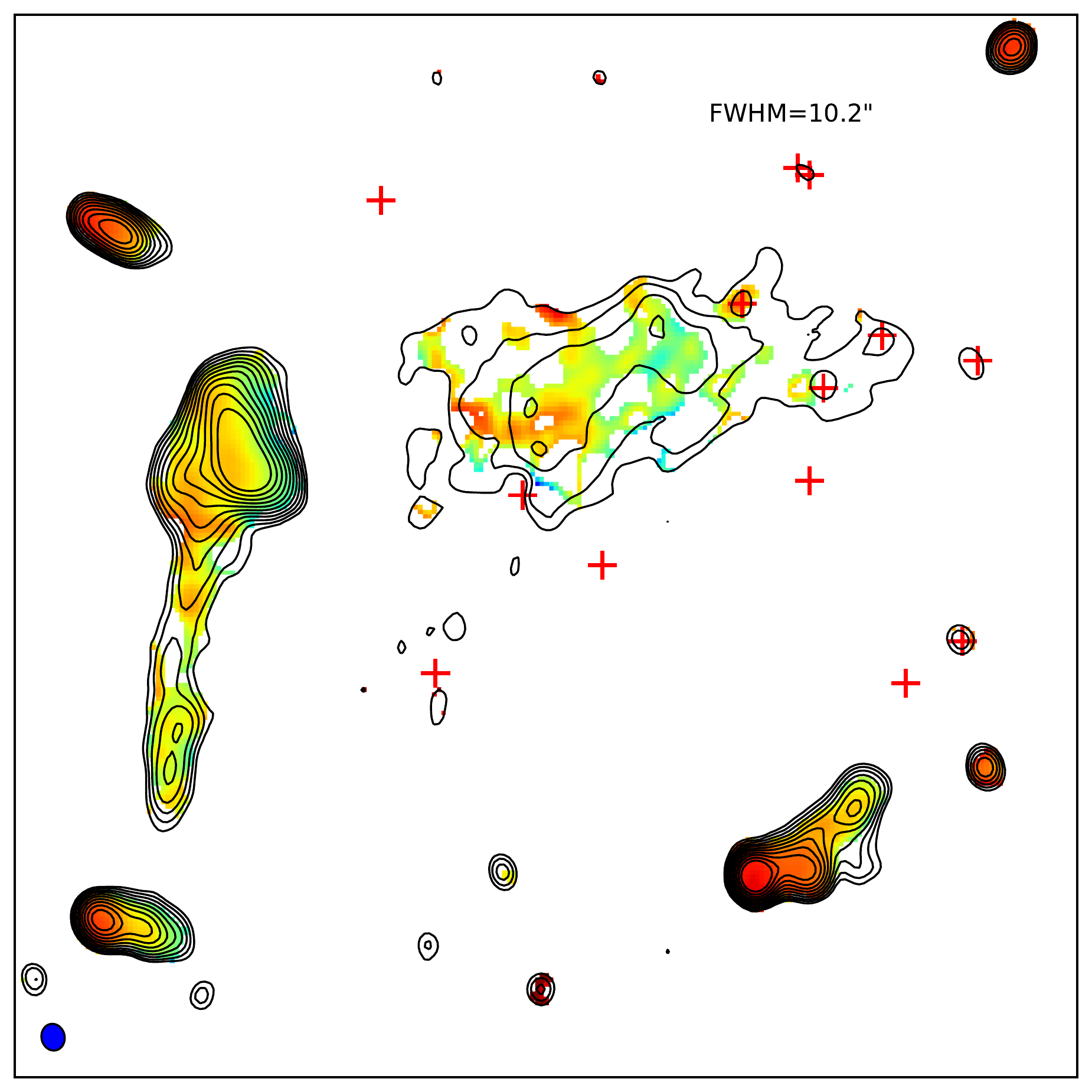}
\includegraphics[height=5.2cm]{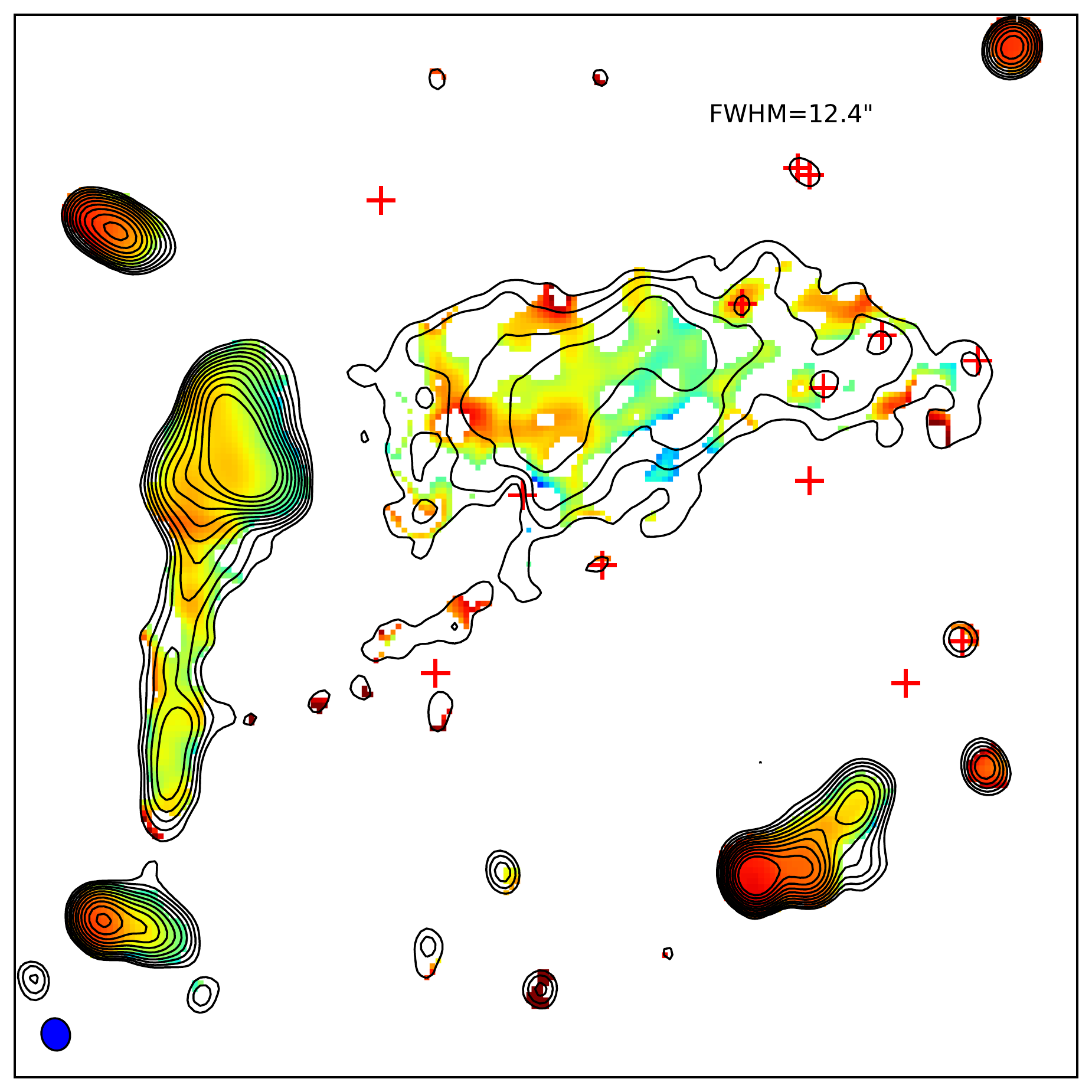}
\includegraphics[height=5.2cm]{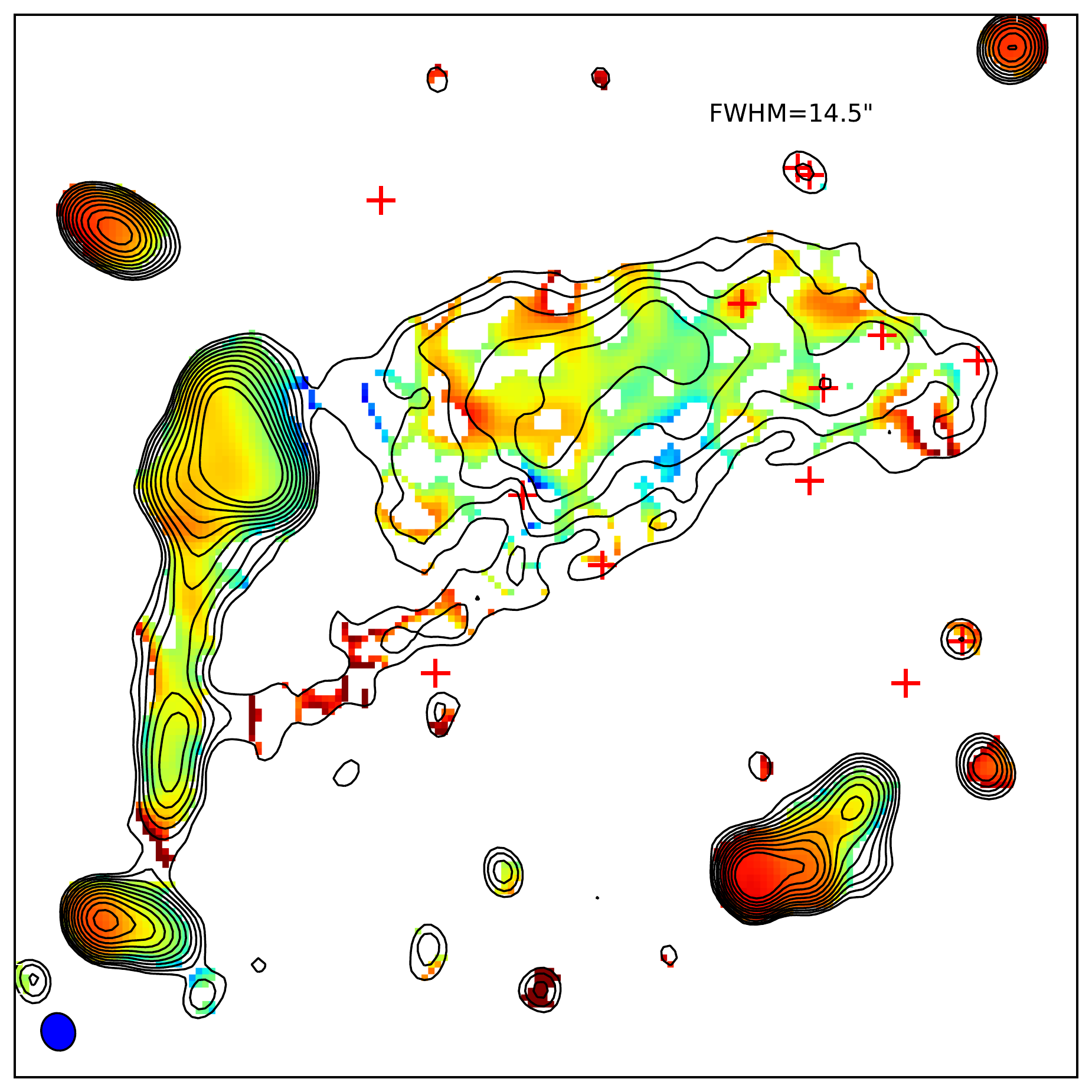}
\includegraphics[height=5.2cm]{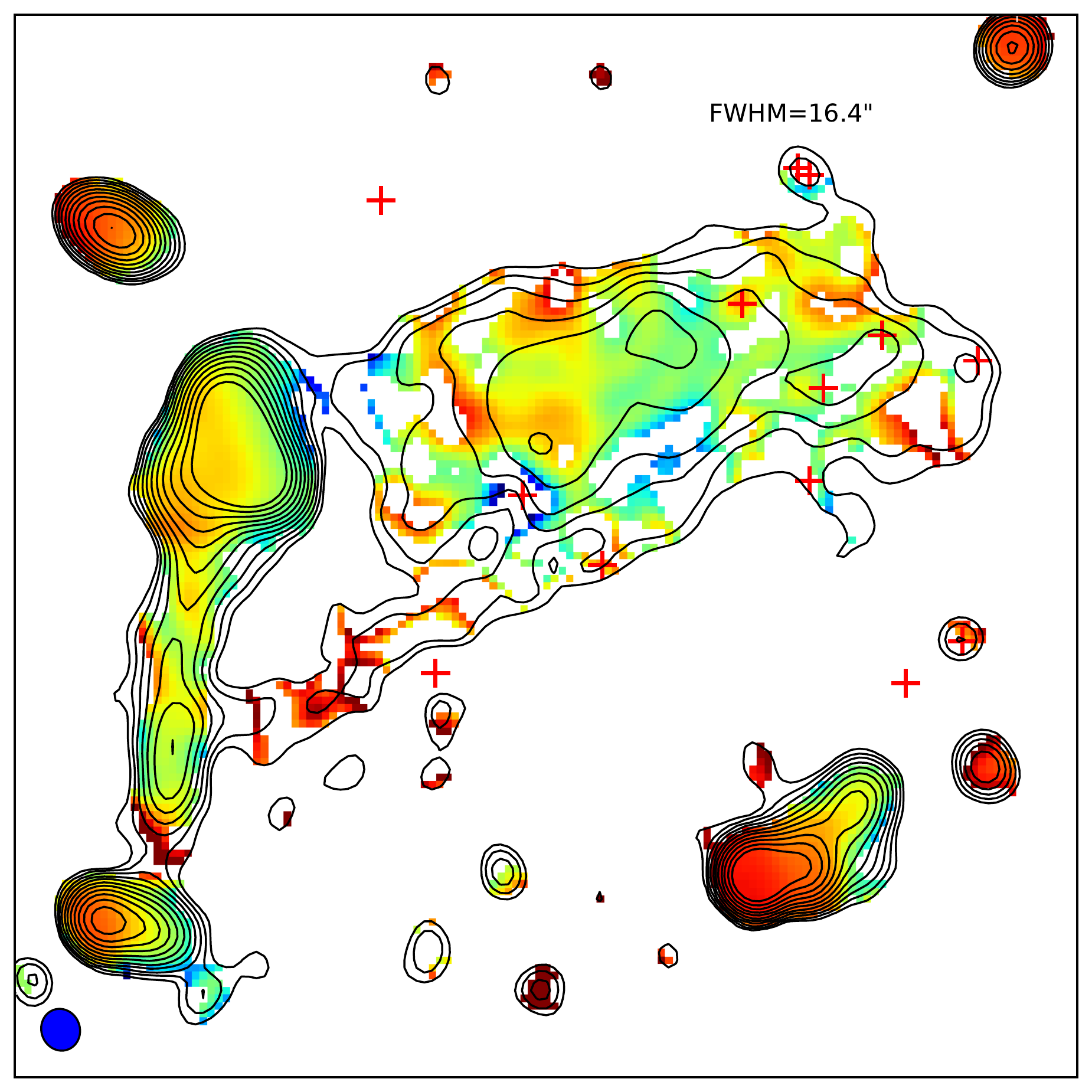}
\includegraphics[height=5.2cm]{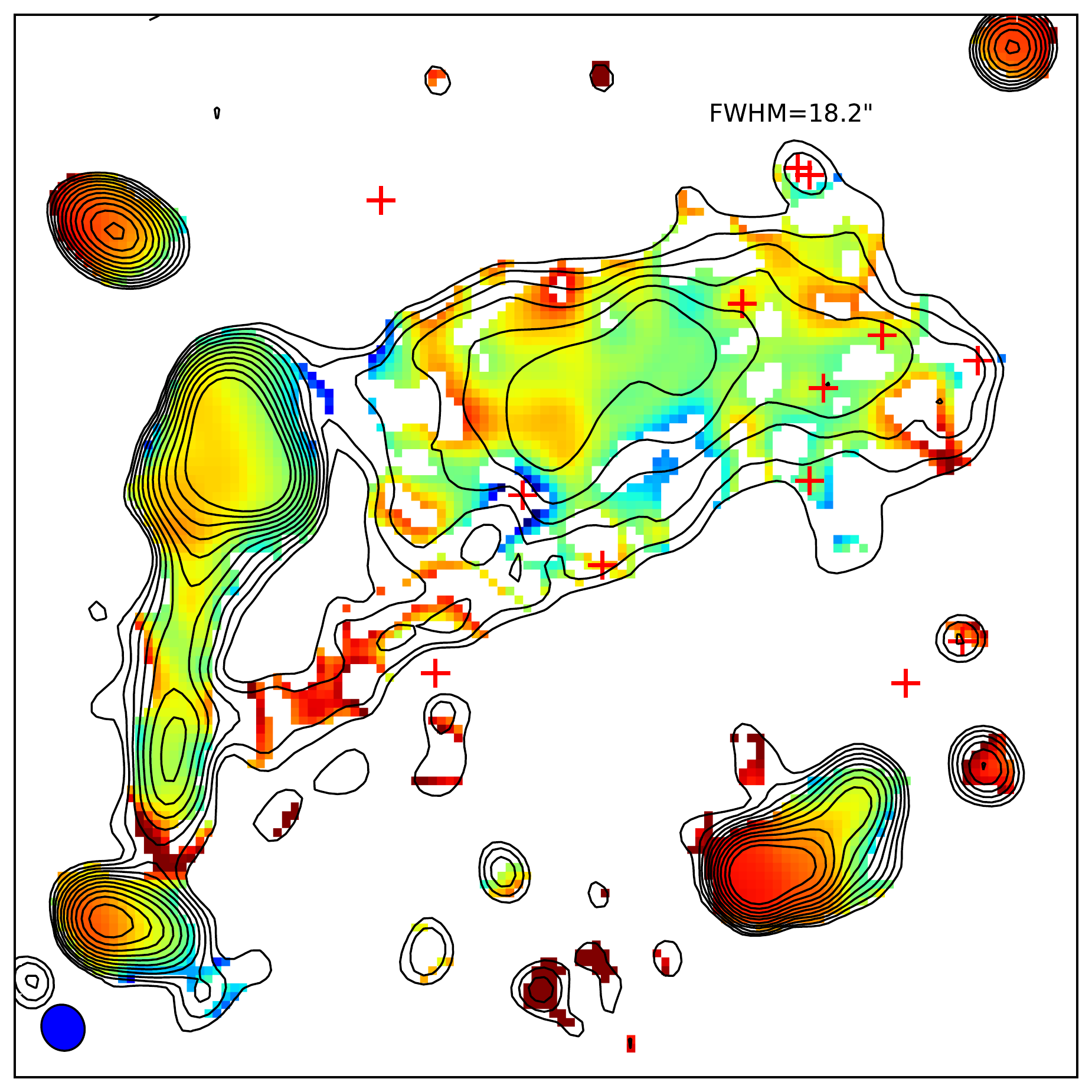}
\includegraphics[height=5.2cm]{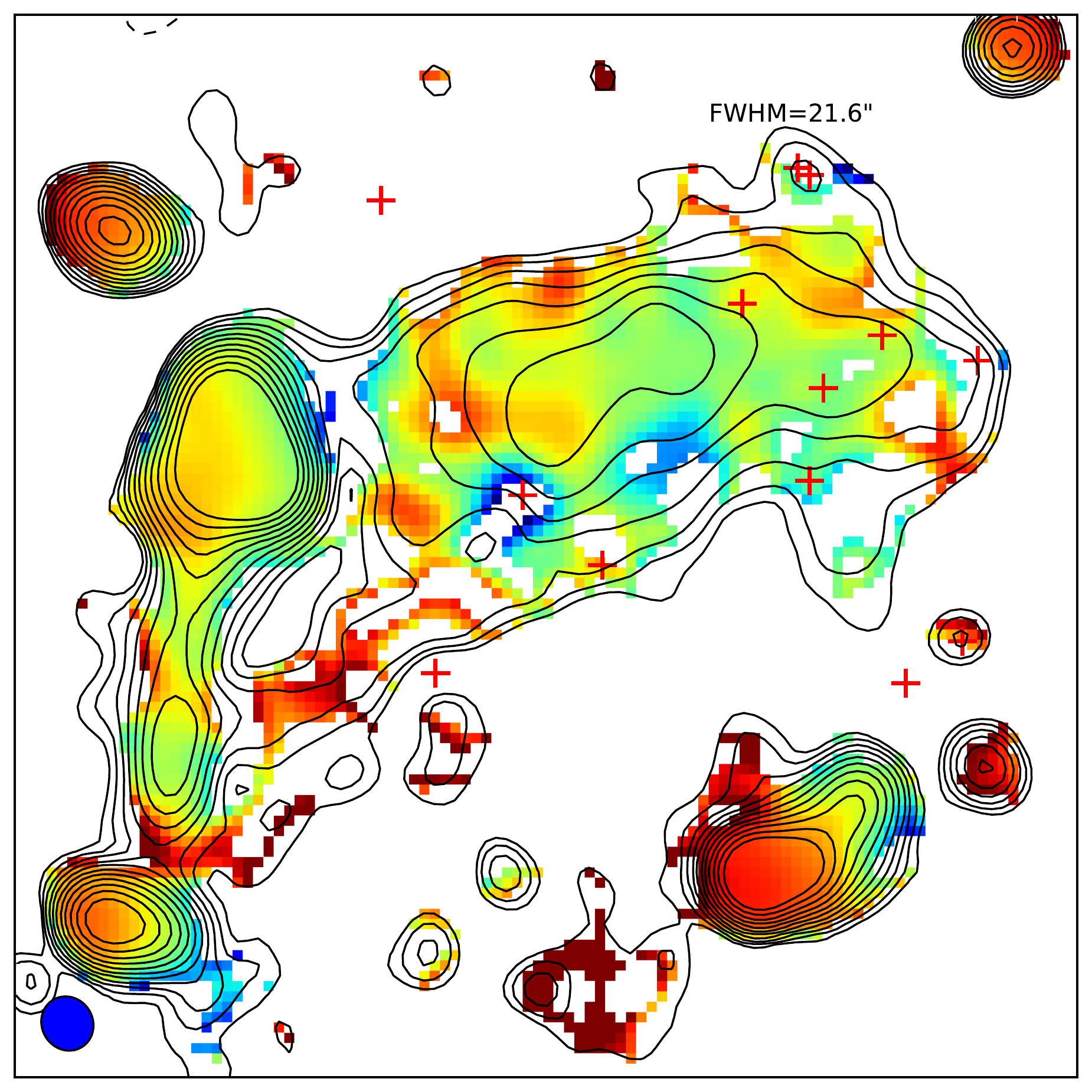}
   \caption{The same as the top panel of Figure \ref{fig-bullet-cluster} but for images created at resolutions ranging from 3.8$\arcsec$ to 21.6$\arcsec$. The synthesised beam is shown in the bottom left corner of the images and its FWHM is displayed on the image. The axis labels, colour scale and contour levels are identical to those in Figure \ref{fig-bullet-cluster}. The spectral index images are blanked for regions where the error exceeds 0.3.}
 \label{fig-bullet-cluster-resolution}
\end{figure*}

The radio halo associated with 1E 0657-55.8 was discovered and well characterised by \cite{Liang_2000}, but our 1.1-3.1\,GHz data are deeper. The 1.1-3.1\,GHz low-resolution continuum and spectral properties of the halo (post source-subtraction) and surrounding region are presented in Figure \ref{fig-bullet-cluster}, higher resolution images are shown in Figure \ref{fig-bullet-cluster-resolution}. In the high resolution images the halo flux per synthesised beam is low but the structure of smaller angular scale sources can be examined. As the resolution is decreased the halo flux per beam increases and the emission from the halo is visible over a large angular scale. We note that the subtraction of contaminating sources, particularly source M, is not perfect and artefacts remain on the image. In all source subtracted images that we present, we show the locations of the subtracted sources so that these regions can be interpreted with caution.

In comparison to the \cite{Liang_2000} continuum images of the halo (Figure 5 of that paper), our detection is at higher significance and the morphology of the halo is more precisely characterised, with sharp edges to the radio halo emission rather than low signal-to-noise poorly defined edges that hinder accurate characterisation and comparison to other datasets. Nevertheless, the \cite{Liang_2000} observations did reveal that the halo was extended along the merging axis, in agreement with our observations of a $\approx$130$\arcsec$ extension in the north-south direction compared to the $\approx$270$\arcsec$ extension in the east-west direction. A key morphological difference is the distinctive radio halo edge on the western side of the halo that we see in Figure \ref{fig-bullet-cluster} which was not apparent in previous data.

\begin{figure}
 \includegraphics[height=7.0cm]{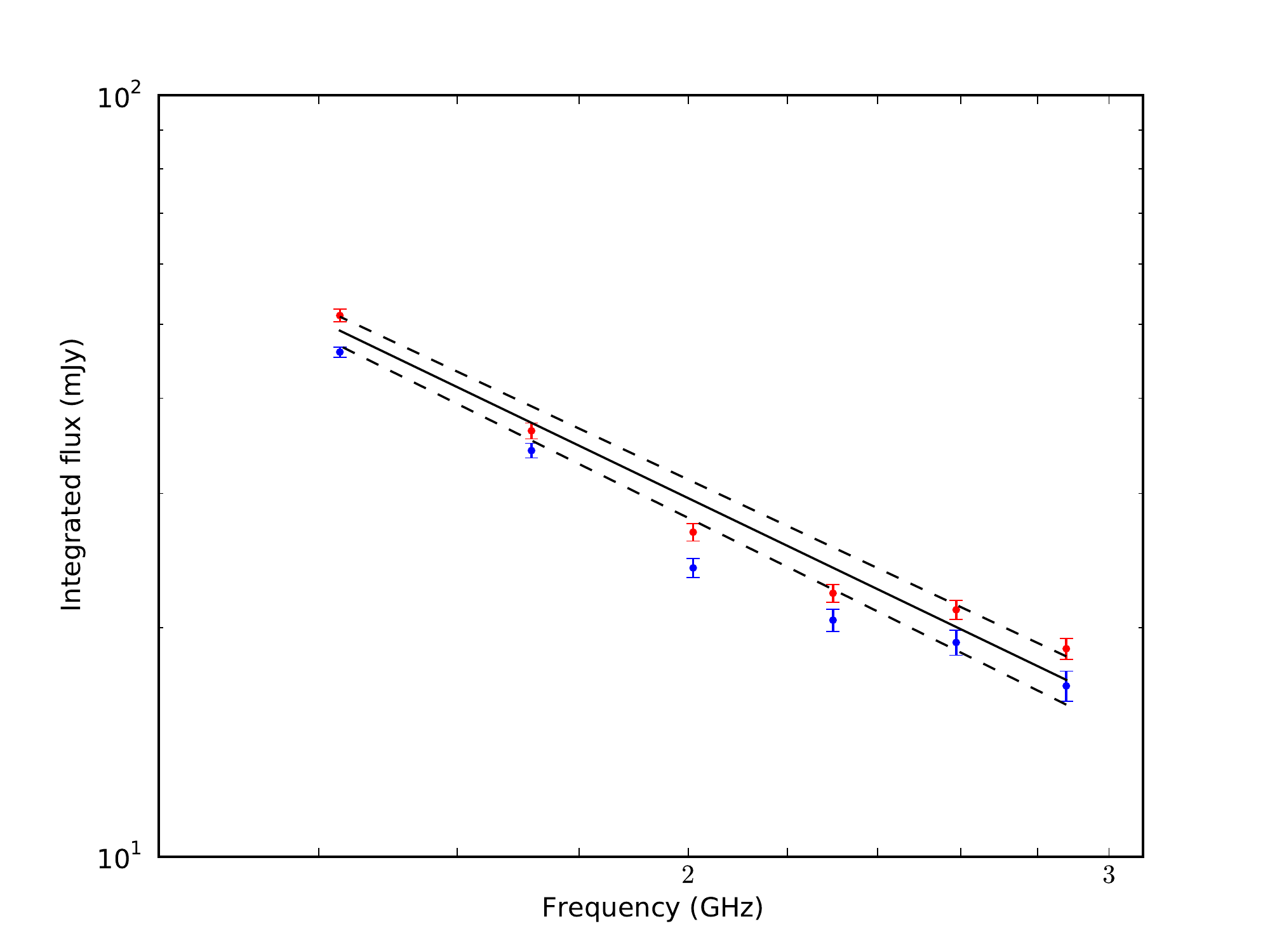} 
   \caption{The frequency dependence of the integrated flux of the radio halo after contaminating sources have been removed. The blue points were measured from the region within the 100$\mu$Jy/beam contour level shown in Figure  \ref{fig-bullet-cluster} that is within the right ascension range 06:58:15 to 06:58:46 and declination range -55:55:00 to -55:58:45 but not within 30$\arcsec$ of source M. The red points were measured from within the \textsc{clean} box region shown in Figure \ref{fig:casa-imagedays_0_1_9_14all-bandrobust-2-deeper_cleanbox}. The solid line shows the best fit to the integrated flux measurement from the larger area and the dashed lines show the errors on this fit.}
    \label{halo-integrated-flux}
\end{figure}

We measure the integrated radio halo flux and its variation with frequency on low resolution narrow-band (see Section \ref{sec:trad-spec}) pre-source subtracted images. We have conducted the measurements over two regions: a) the 100$\mu$Jy/beam contour level shown in Figure  \ref{fig-bullet-cluster} that is within the right ascension range 06:58:15 to 06:58:46 and declination range -55:55:00 to -55:58:45 but not within 30$\arcsec$ of source M (area 8.5 square arcminutes); b) the \textsc{clean} box shown in Figure \ref{fig:casa-imagedays_0_1_9_14all-bandrobust-2-deeper_cleanbox} (area 18.4 square arc minutes). We subtracted the flux of sources within the regions of integration using the parameters in Table \ref{tab:ATCA_SOURCES} (for sources where we were unable to measure the spectral properties we use the 1.1-3.1\,GHz flux to approximate the 2.1\,GHz flux and assume $\alpha^{3.1}_{1.1}=-0.7\pm0.5$) and propagate the errors into our measurements of the integrated radio halo flux.  From the smaller area we find an integrated 2.1\,GHz flux of 24.7$\pm$1.5\,mJy and $\alpha^{3.1}_{1.1}=-1.57\pm0.05$ and from the larger area we find an integrated 2.1\,GHz flux of 27.5$\pm$1.7\,mJy and $\alpha^{3.1}_{1.1}=-1.50\pm0.04$. Our measurements of the integrated halo properties are presented in Figure \ref{halo-integrated-flux}.

\begin{figure}
 \includegraphics[height=7.0cm]{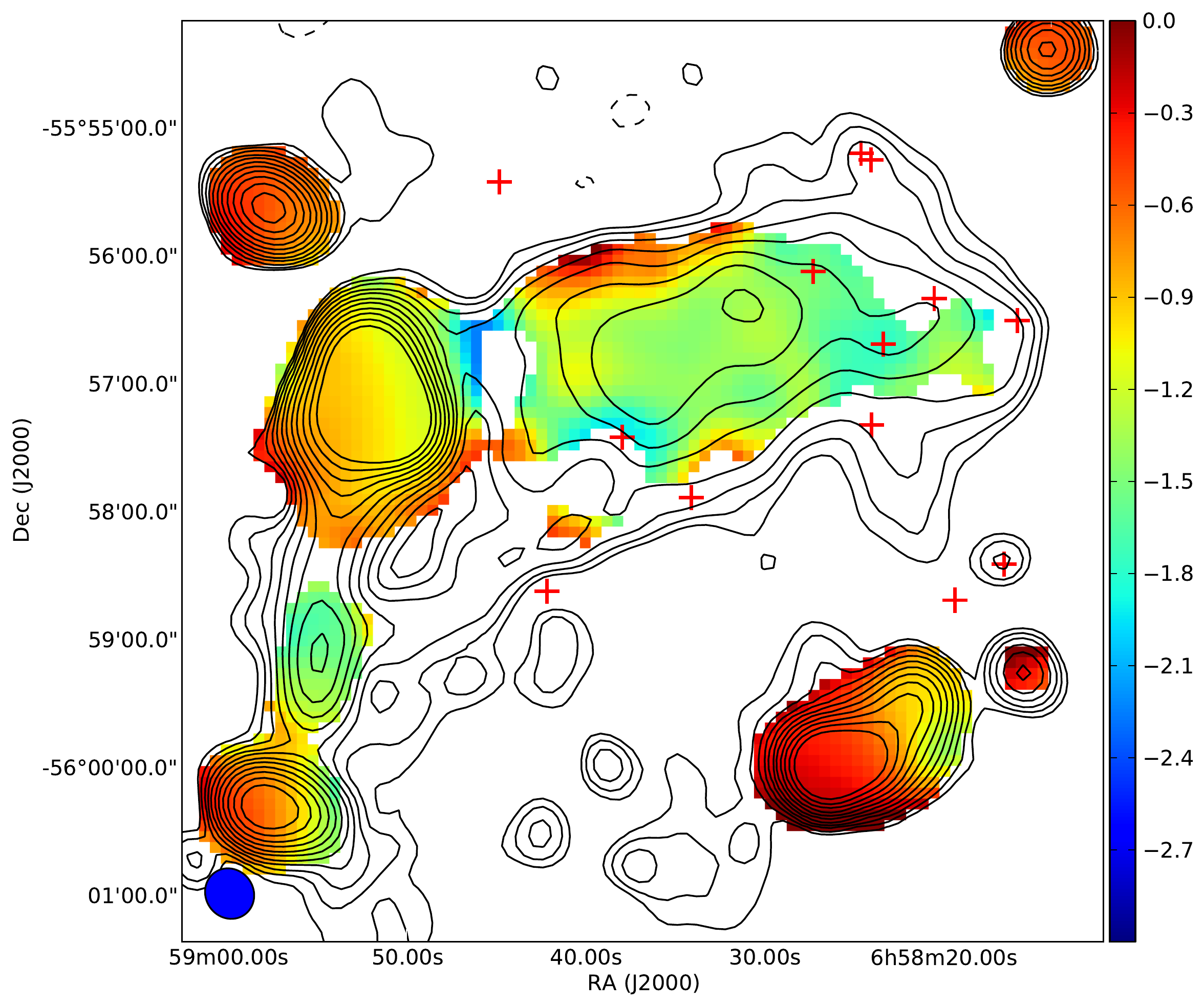} 
   \caption{A low resolution (synthesised beam FWHM equals 28.6$\arcsec$) spectral index map created from the narrow-band method described in Section \ref{sec:trad-spec}.  All pixels with errors greater than 0.3 are blanked. In regions where we were able to determine the spectral index this spectral index map is consistent with the spectral index map created using the wide-band method (see Section \ref{sec:rau_method}) which shown in Figure \ref{fig-bullet-cluster}. The contour levels are identical to those in Figure \ref{fig-bullet-cluster}. }
    \label{traditional-spec}
\end{figure}

\begin{figure}
\centering
 \includegraphics[height=7cm]{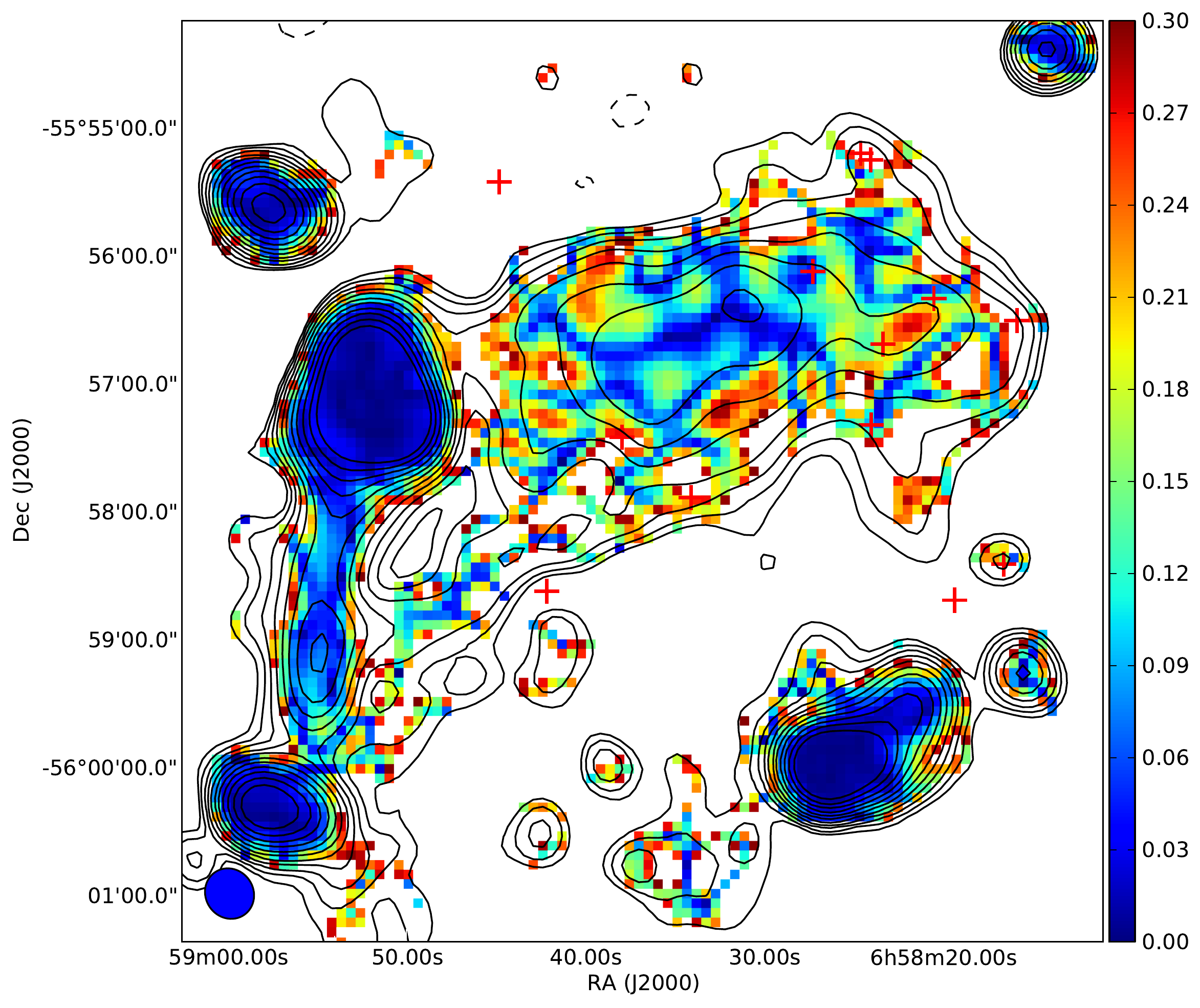}
\caption{The spectral index error image for the 1.1-3.1\,GHz spectral index map in Figure \ref{fig-bullet-cluster} that was created using the wide-band method described in Section \ref{sec:rau_method}. The maximum spectral index error is 0.3. The contour levels are identical to those in Figure \ref{fig-bullet-cluster}. }
 \label{fig-bullet-cluster-error}
\end{figure}

We have measured the spectral index across the halo in two ways (see Section \ref{sec:spectral_index}); spectral index images created at low resolution using the wide-band and narrow-band methods are shown in Figures \ref{fig-bullet-cluster} and \ref{traditional-spec} respectively (the wide-band spectral index error image is shown in Figure \ref{fig-bullet-cluster-error}). The two spectral index methods produce consistent results and are in agreement with our derived spectral index of the integrated halo flux. The spectral index images reveal that the central region of the halo has a spectral index of $\alpha^{3.1}_{1.1} \approx -1.4$ and that there are variations in the spectral index across the halo.

\begin{figure*}
 \includegraphics[height=7.0cm]{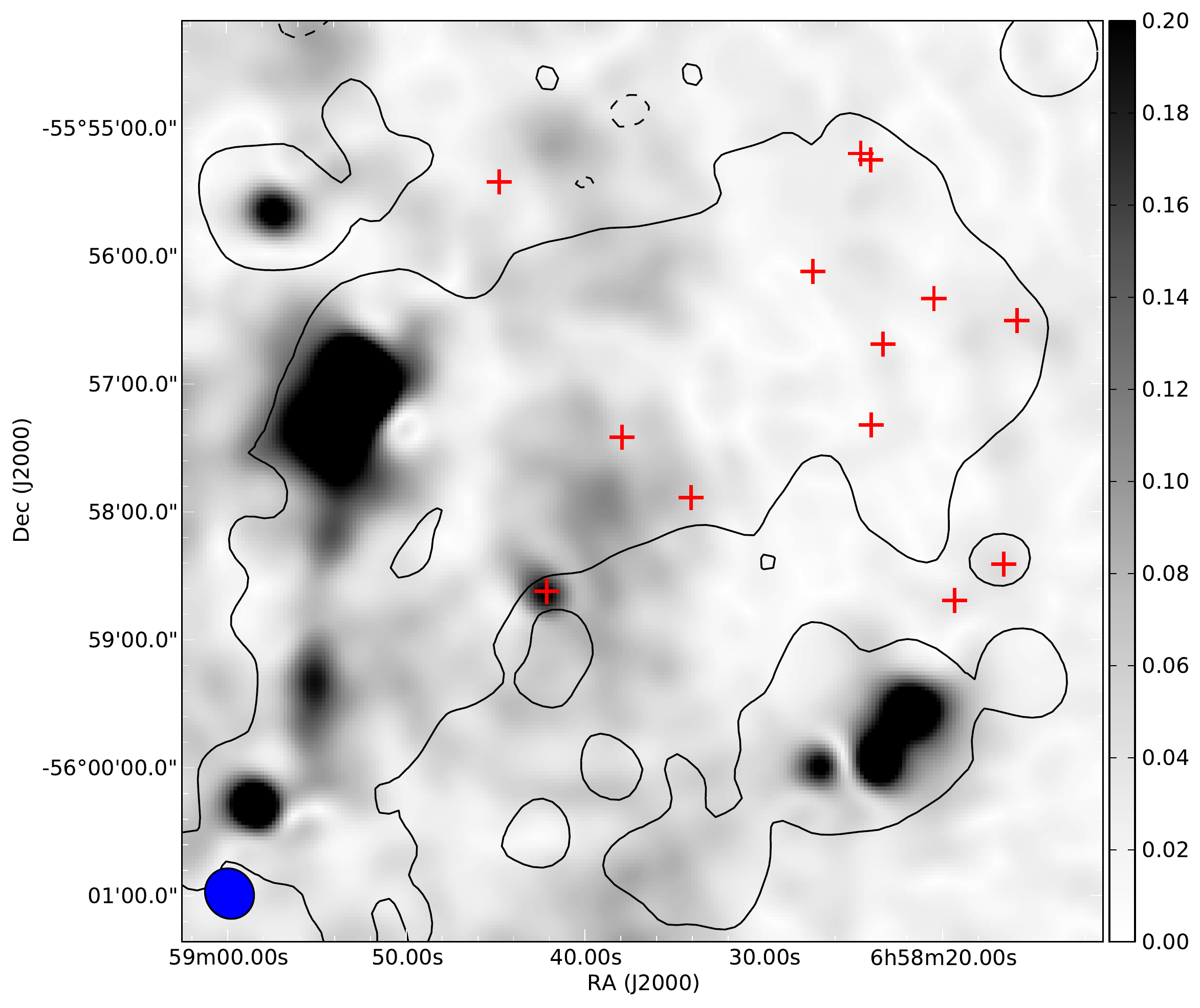}  \includegraphics[height=7.0cm]{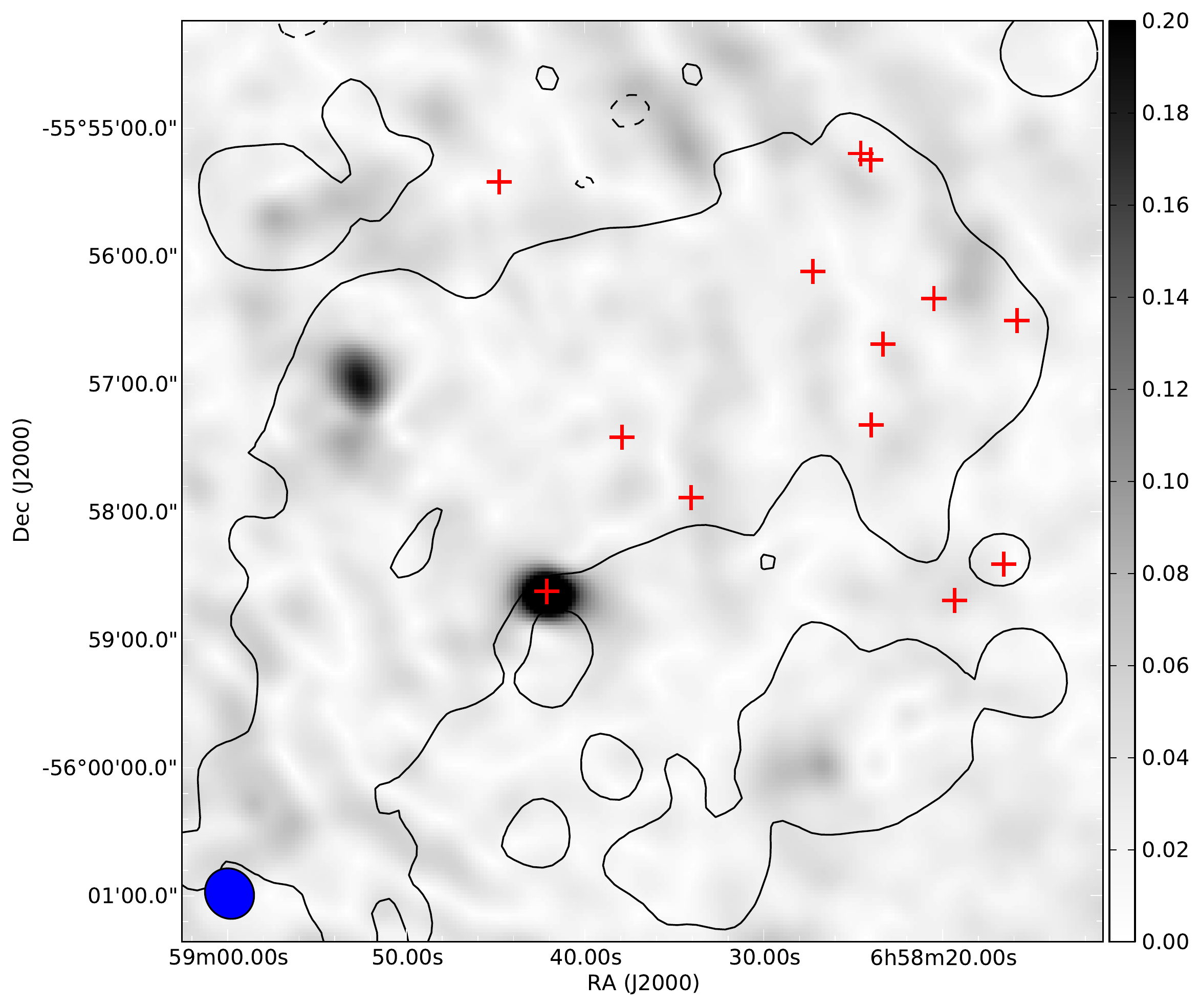} 
   \caption{Left: polarised intensity image at RM=$-30$\,rad/m$^{2}$ where $\sigma_{Q,U}=31\mu$Jy/beam. Right:  polarised intensity image at RM=$-236$\,rad/m$^{2}$ where $\sigma_{Q,U}=23\mu$Jy/beam. The polarised intensity images are at low resolution (FWHM=21$\arcsec$) and the greyscale is in mJy beam$^{-1}$ RMSF$^{-1}$. The contour shows the 5$\times$25$\mu$Jy/beam level of the source subtracted Stokes I image at 23.3$\arcsec$ resolution.}
    \label{polarised_intensity}
\end{figure*}

\begin{figure}
 \includegraphics[height=7.0cm]{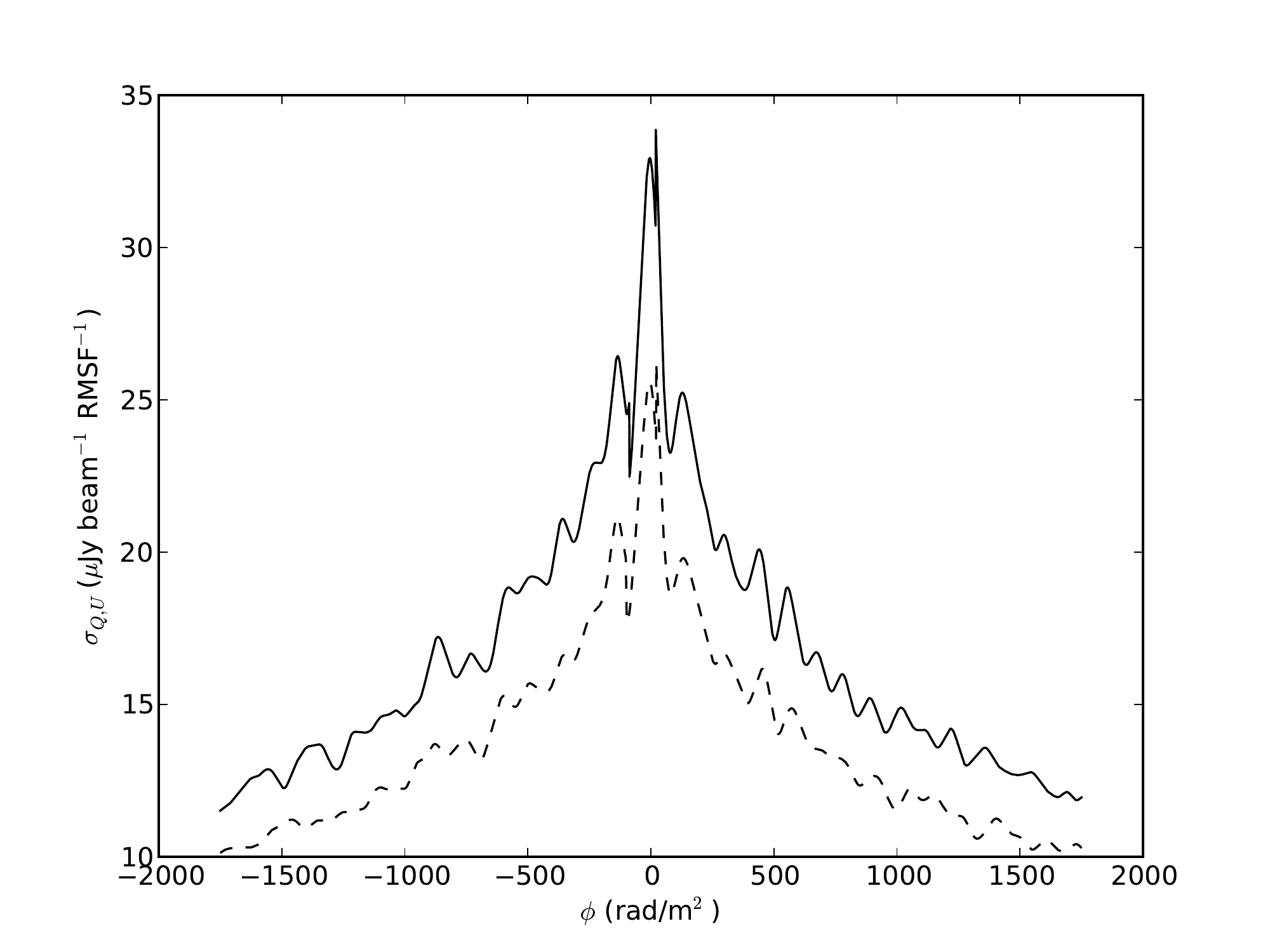}  
   \caption{The noise ($\sigma_{Q,U}$) in F($\phi$) as a function of $\phi$. The dashed line shows the noise when the Q and U images were made with a resolution of 17$\arcsec$ and the solid line shows the noise on the 21$\arcsec$ resolution images. The noise is highest at a wide range of $\phi$ around $\phi$=0. In this region there is more emission from polarised sources in the field.}
    \label{rm-cube-noise}
\end{figure}

We have used rotation measure synthesis to search the wide-band dataset for polarised emission. We created RM-cubes from the 1.1-3.1\,GHz dataset (see Section \ref{sec:rm-cube}) at resolutions of 17$\arcsec$ and $21\arcsec$, Figure \ref{polarised_intensity} shows several planes of the low resolution RM-cube. In the region of the halo the standard deviation of the Stokes Q and U components, $\sigma_{Q,U}$, of the complex RM-cube varied with RM significantly (see Figure \ref{rm-cube-noise}), with lower noise levels obtained at large positive or negative RM away from all the signal in the Q and U images. We searched through the RM-cubes in the region of the radio halo (which we defined as right ascension 06:58:15 to 06:58:45 and declination -55:55:00 to -55:58:00) for polarised emission that exceeded 5$\times \sigma_{Q,U}$, but were unable to detect any emission. The peak intensity of the 1.1-3.1GHz Stokes I images is 1.3\,mJy/beam at 21$\arcsec$ and 0.8\,mJy/beam at 17$\arcsec$, while the maximum 5$\sigma_{Q,U}$  of the F($\phi$) spectrum at these resolutions is 170$\mu$Jy/beam RMSF$^{-1}$ and 130$\mu$Jy/beam RMSF$^{-1}$, respectively. The non-detection of polarised emission implies that in the region of peak radio emission, the fractional polarisation in the 1.1-3.1\,GHz images is less than 13\% when observed with 21$\arcsec$ resolution and less than 16\% at $17\arcsec$ resolution.

To calculate a limit on fractional polarisation at 1.4\,GHz we used the narrow-band polarisation technique described in Section \ref{sec:pol_freq}. For a 20$\arcsec$ resolution polarisation image of a 100\,MHz wide band centred at 1.48\,GHz, in which $\sigma_{Q,U}$ is 54$\mu$Jy/beam RMSF$^{-1}$, we found no polarised emission at greater than 5$\sigma_{Q,U}$ within the region of the radio halo. From our narrow-band images we estimate a radio halo peak flux of 3.1\,mJy/beam at 1.48\,GHz and therefore our 5$\sigma_{Q,U}$ upper limit on the fractional polarised intensity at this frequency is 9\%.

\subsection{LEH2001\,J06587-5558}

J06587-5558 is a steep spectrum, extremely polarised, extended, extragalactic source observed coincident with the south-east peripheral region of 1E 0657-55.8 (source M in Figure \ref{fig:casa-imagedays_0_1_9_14all-bandrobust-2-deeper_cleanbox} and Table \ref{tab:ATCA_SOURCES}). A study of this source was presented by \cite{Liang_2001}, who concluded that it is either a gravitationally lensed source, a high redshift radio galaxy or a cluster relic. They measured a flux density of 37$\pm2$\,mJy at 1.3\,GHz, spectral indices $\alpha^{1.3}_{2.2}$ = -1.0 and $\alpha^{4.8}_{8.8}$=-1.5, and found that the source rapidly depolarises with increasing wavelength, with integrated fractional polarisations of 54$\pm$11\%, 45$\pm$9\%, 4.3$\pm0.5$\% and $<$0.5\% at 8.8, 4.8, 2.2 and 1.3\,GHz, respectively.

\begin{figure}
 \includegraphics[height=7.0cm]{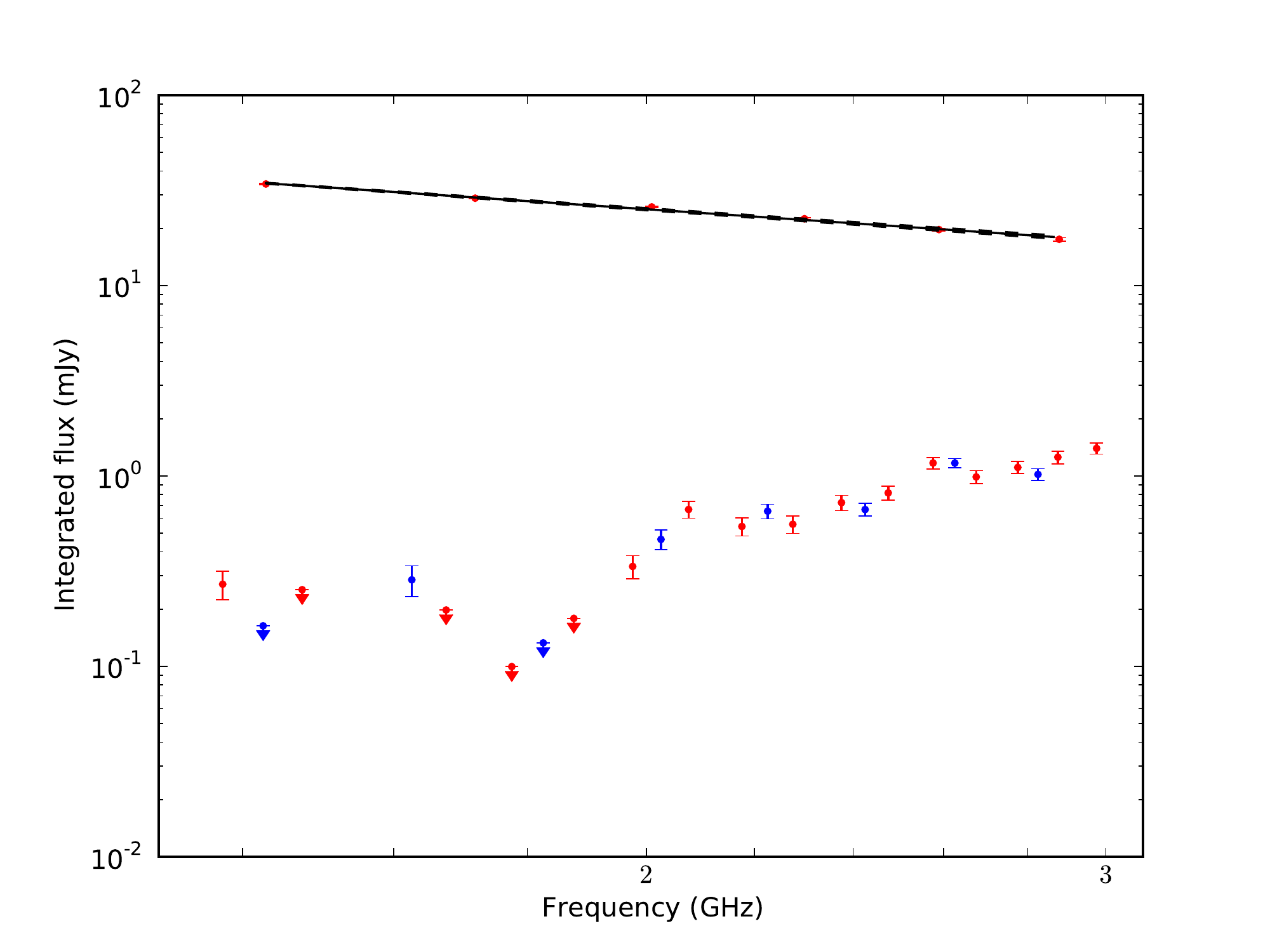} 
   \caption{The total intensity and polarisation spectral energy distributions over 1.1-3.1\,GHz for the source J06587-5558. The red points that have been fitted with a power law (black line) show the Stokes I emission from J06587-5558. The unfitted red points show the polarised emission from J06587-5558 where the Stokes Q and U data were imaged in 100\,MHz sections. The unfitted blue points again show the polarised emission from J06587-5558 but where the Stokes Q and U were imaged in 200\,MHz sections. Upper limits (5$\sigma_{Q,U}$) on the polarised intensity are shown with arrows.}
    \label{fig:liang-properties}
\end{figure}

From high resolution Stokes I images of J06587-5558 we measure a 1.4\,GHz flux of 35.1$\pm$0.5\,mJy and $\alpha^{1.1}_{3.1}$ =-0.93$\pm0.01$, both in agreement with the results of \cite{Liang_2001}. We find no evidence for a change in spectral index over the observed frequency range (see Figure \ref{fig:liang-properties}) and combining this result with the measurements of \cite{Liang_2001} indicates that a break in the spectral index must occur between 3.0\,GHz and 4.8\,GHz. For the polarisation analysis, we split the 1.1-3.1\,GHz band into 100\,MHz sub-bands and follow the procedure outlined in Section \ref{sec:pol_freq} to determine the frequency dependence of the polarised emission from this source. We perform the same analysis but with 200\,MHz sub-bands and find that the polarised emission in these wider sub-bands is consistent with measurements from the 100\,MHz analysis, indicating that for this analysis we are not significantly effected by bandwidth depolarisation. The measured polarisation and Stokes I flux for this source are given in Figure \ref{fig:liang-properties}. In agreement with \cite{Liang_2001}, we find that the source depolarises at lower frequency. Our measurements of polarised emission at frequencies less than 2.0\,GHz are at low significance but we find that at 1.38\,GHz the fractional polarisation is 0.8$\pm$0.1\%. At 2.2\,GHz we measure a fractional polarisation of 2.7$\pm0.3$\% and this rises to 7.0$\pm0.1$\% at 2.9\,GHz. Our 2.2\,GHz measurement of the fractional polarisation is a little lower than was determined by \cite{Liang_2001} and further measurements would be useful to reconcile the difference.

\begin{figure}
 \includegraphics[height=7.0cm]{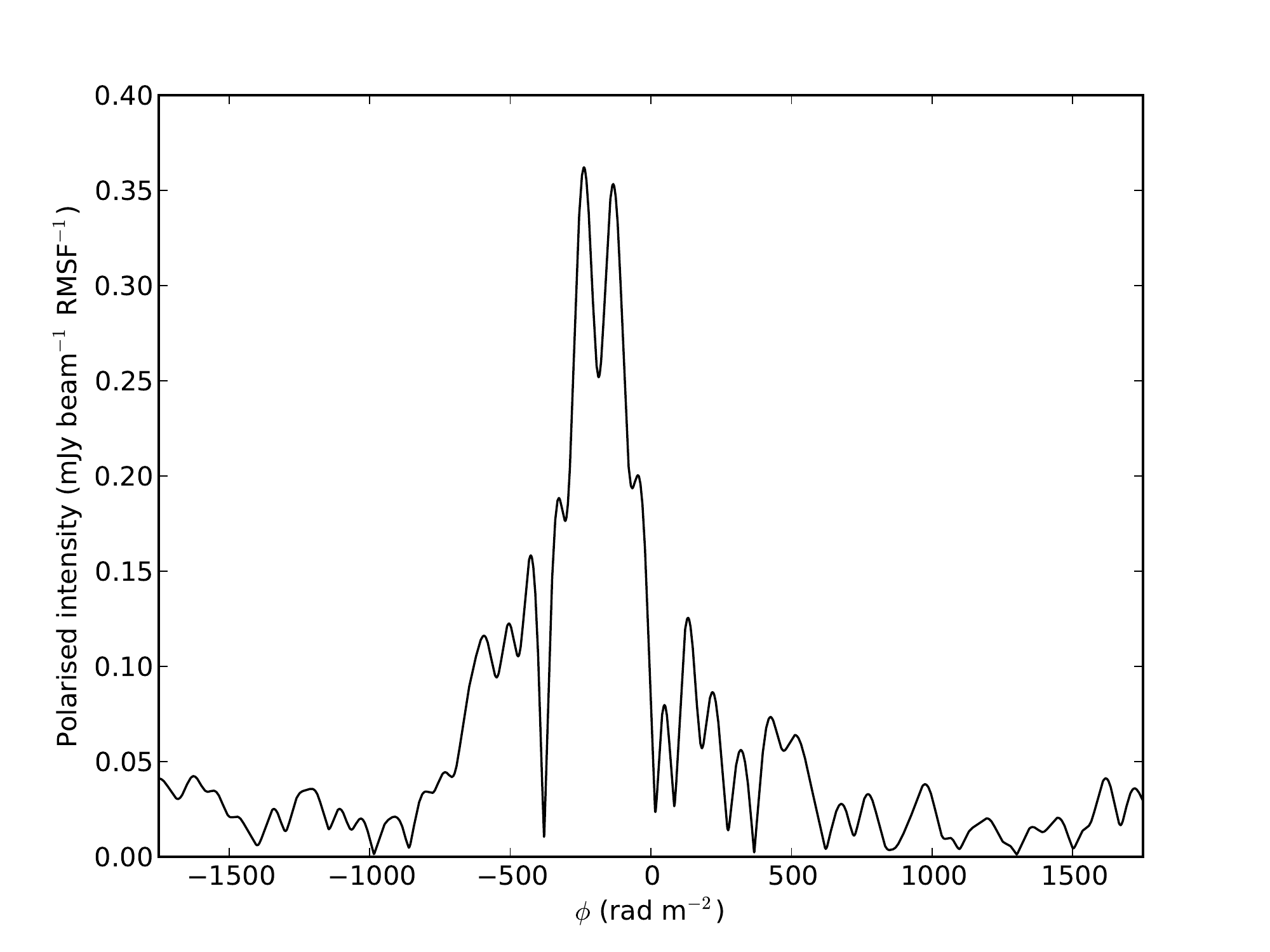} 
   \caption{The F($\phi$) spectrum at 6:58:42.1 -55:58:37.6, the location of the Stokes I peak of J06587-5558.}
    \label{fig:phi-spectrum-liang}
\end{figure}

The Galactic magnetic field in the direction of the cluster is $+4.8\pm44.0$\,rad/m$^{2}$ (\citealt{Oppermann_2012}) but the F($\phi$) spectrum of this source reveals that the polarised intensity peaks at a $\phi= -238$\,rad/m$^{2}$. This is in agreement with the  \cite{Liang_2001} measured value of -266$\pm$37\,rad/m$^{2}$ but, as shown in Figure \ref{fig:phi-spectrum-liang}, the F($\phi$) spectrum of this source likely has multiple components.

\section{Discussion}

In Figure \ref{fig-bullet-cluster}, we compare our low-resolution radio halo image to the observed X-ray brightness (\citealt{Markevitch_2006}) and to the weak-lensing mass reconstruction (\citealt{Clowe_2006}). These images reveal that the morphology of the radio halo and thus the population of synchrotron emitting electrons has similarities to, but does not strictly follow, the X-ray observed bremstrahlung emission that traces the intra-cluster medium (see also Figure \ref{1d-slices}), nor does it trace the distribution of galaxies that extends well beyond the western edge of the radio halo. 

Unlike for the Coma cluster (\citealt{Giovannini_1993}), we do not observe a radial steepening of the spectral index and instead observe that there are localised regions within the spectral index image that are different from the average (see Figures \ref{fig-bullet-cluster}, \ref{fig-bullet-cluster-resolution} and \ref{traditional-spec}). For example, on the eastern edge of the radio halo (06:58:40 -55:57:00) close to the proposed radio relic, the spectral index is flatter than average; this flattening may be associated with the shock that created the proposed relic or a counter shock of the prominent western bow shock. Alternatively, the lack of an ordered spectral index distribution may be because the halo is young and in formation, with turbulence associated with the ongoing merger still altering its structure.

\subsection{Integrated halo properties}

\cite{Liang_2000} measured an integrated radio halo flux of 78$\pm$5\,mJy at 1.3\,GHz from a large region (32.7 square arcminutes) around 1E 0657-55.8. They also measured an integrated 1.3\,GHz flux of $\approx$50\,mJy from the 7.3 square arcminute region of bright radio halo emission (see Figure 7. of \citealt{Liang_2000}). Similarly we have made two measurements of the integrated flux: one over a small area (8.5 square arcminutes), which is entirely filled with significant radio halo emission; and another over a large area (18.4 square arcminutes), which encompasses the halo and surrounding region. From the fit to our measurements of the integrated radio halo flux as a function of frequency we determine 2.1\,GHz (1.3\,GHz) integrated flux densities of 24.7$\pm$1.5\,mJy (52.5$\pm$2.1\,mJy) and 27.5$\pm$1.7\,mJy (56.4$\pm$2.3\,mJy) from the smaller and larger regions respectively (see Figure \ref{halo-integrated-flux}). Our measurements are in good agreement with the small area measurements by \cite{Liang_2000}. However, a discrepancy arises when the area integrated over is much larger than the region of significant radio emission. Our measurement remains approximately constant with integration area whereas the \cite{Liang_2000} measured value rises to 78$\pm$5\,mJy at 1.3\,GHz. Our spectral index measurements ($\alpha^{3.1}_{1.1}=-1.57\pm0.05$ and $\alpha^{3.1}_{1.1}=-1.50\pm0.04$) are steeper than the \cite{Liang_2000} measurement of $\alpha^{8.8}_{0.8}=-1.3\pm0.1$.

Future observations will be necessary to reconcile the difference between our large area integrated flux estimates and those by \cite{Liang_2000} but we note that our flux and spectral index measurements of less complex objects, such as LEH2001\,J06587-5558, are in good agreement with the \cite{Liang_2000} measurements.

\subsection{Radio vs X-ray brightness}

\begin{figure*}
\centering
     \includegraphics[width=5.8cm]{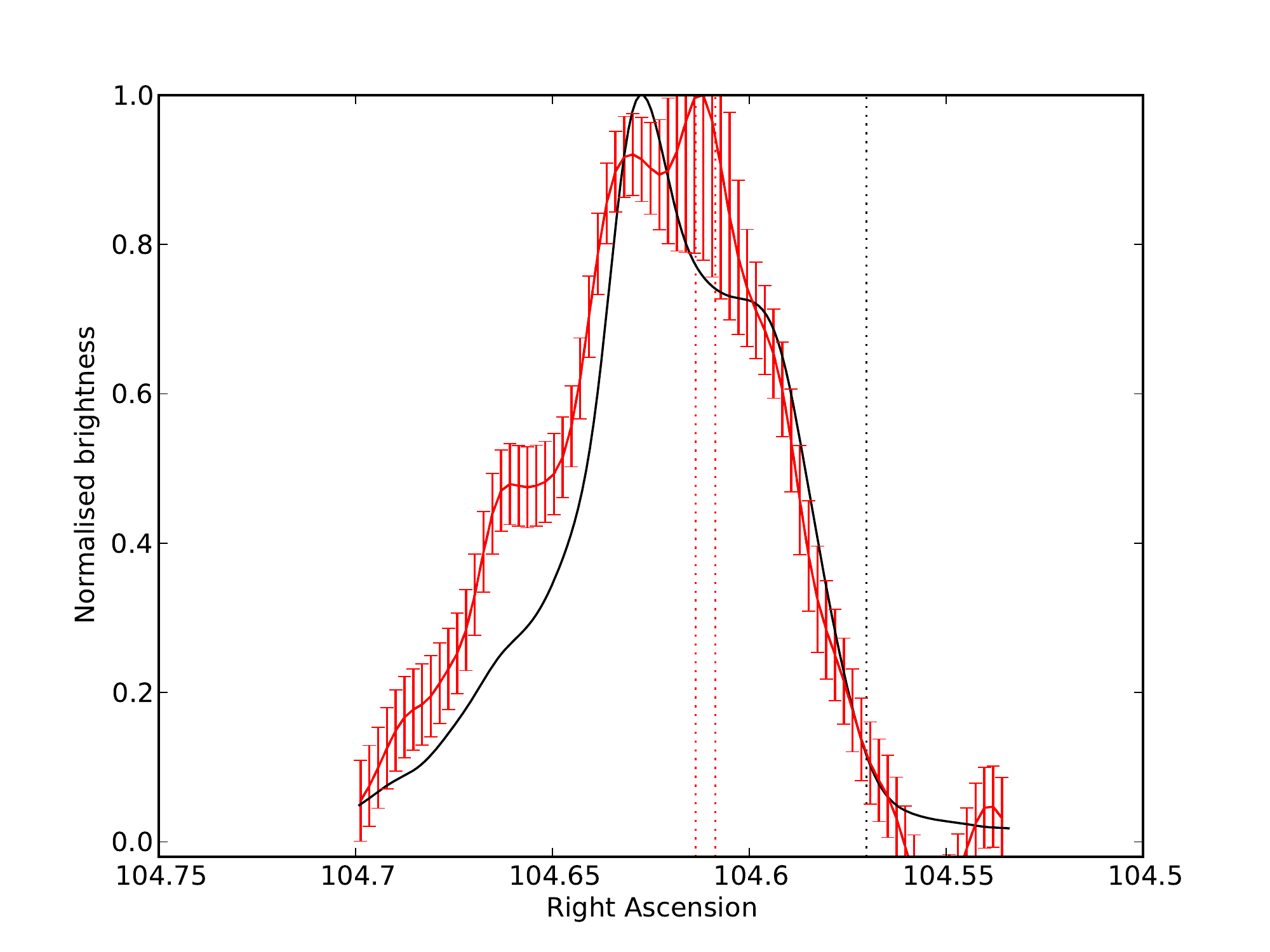} \includegraphics[width=5.8cm]{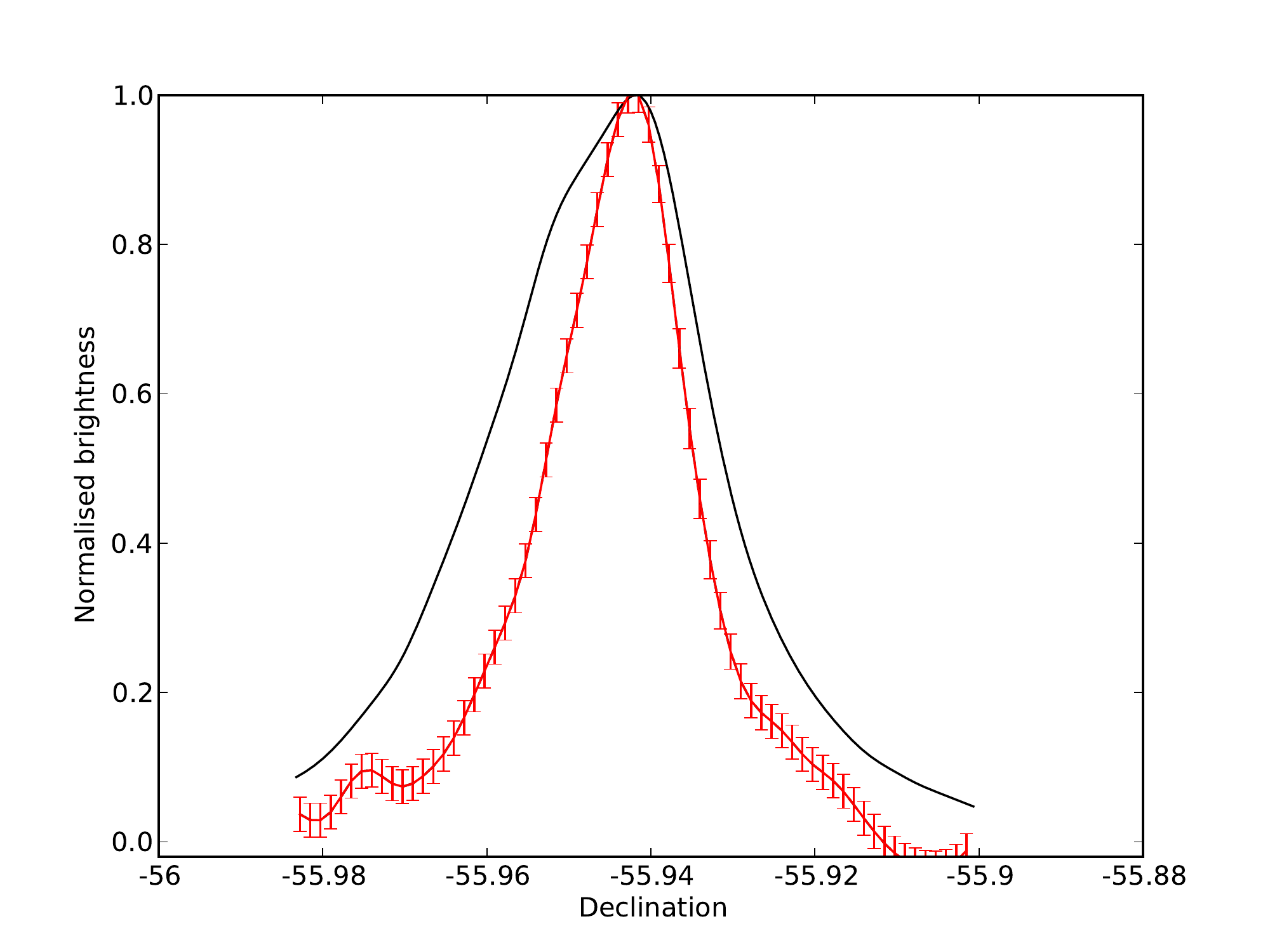} 
     
     Left: a slice through right ascension at a declination of -55:56:00.0. Right: a slice through declination at right ascension 06:58:30.0. The slice through declination also passes close (3$\arcsec$) to the peak in the centroid of the X-ray emission from the main cluster at 55.94$^\circ$.
          
   \includegraphics[width=5.8cm]{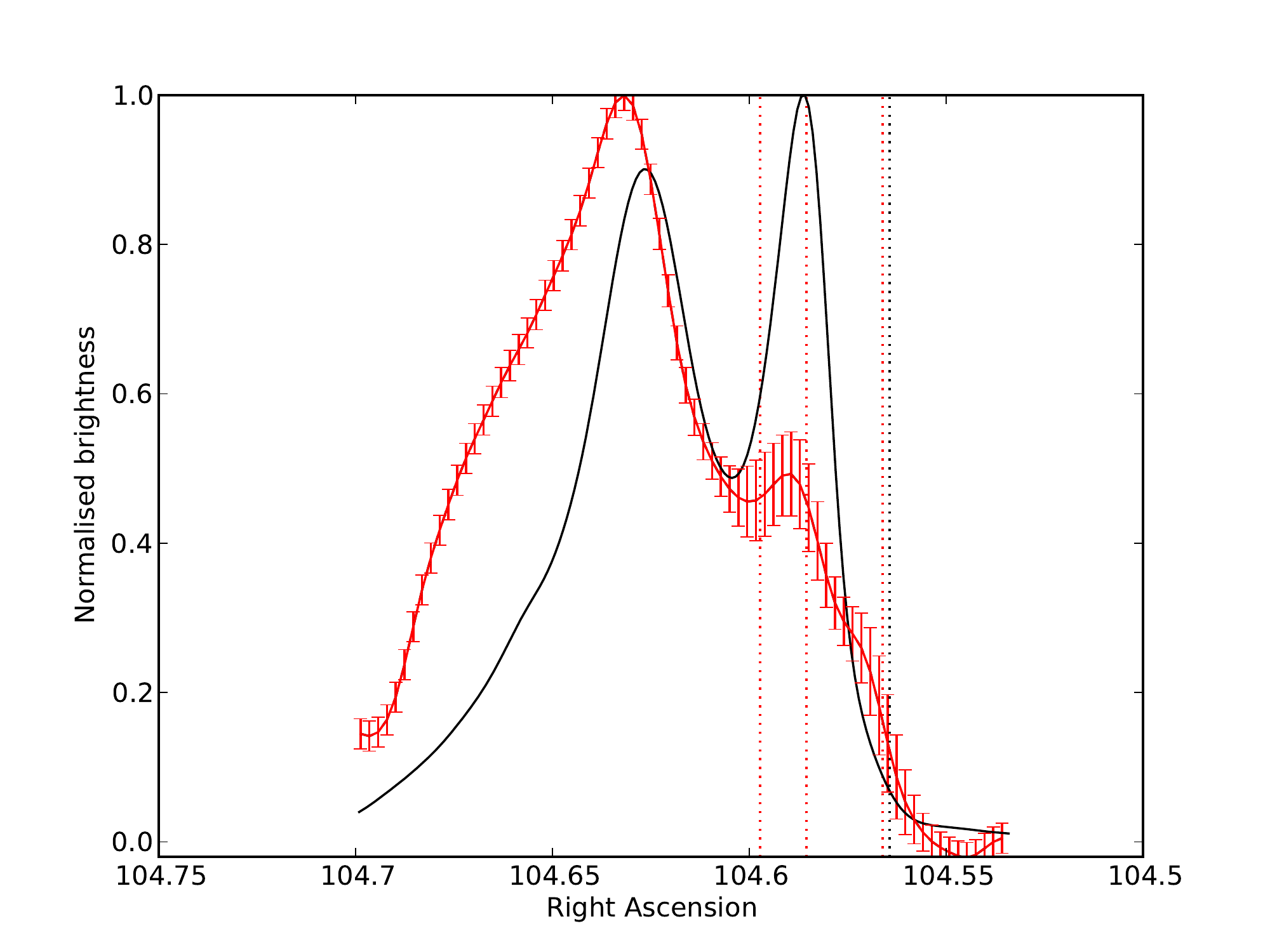} \includegraphics[width=5.8cm]{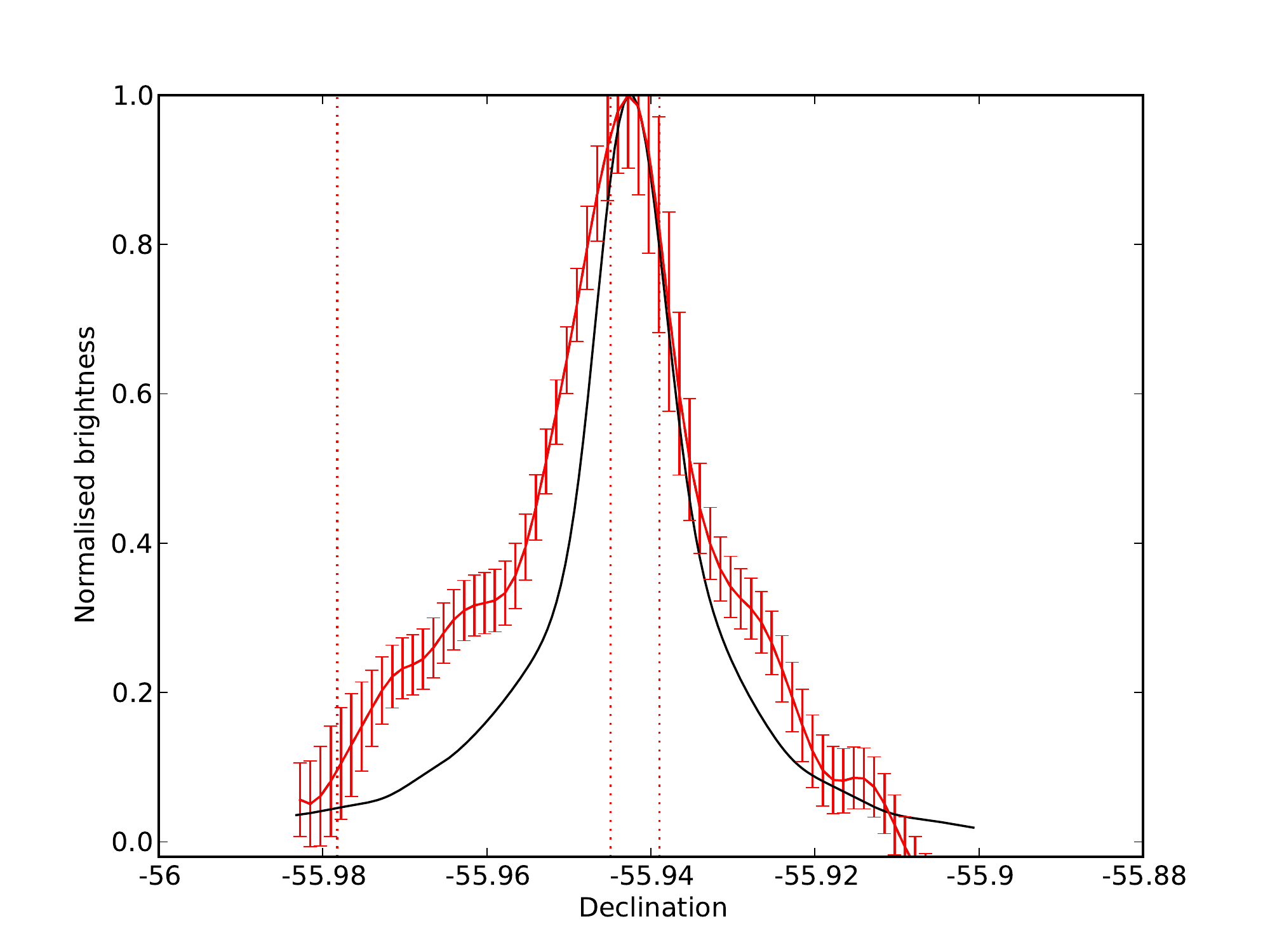}    
   
   Left: a slice through right ascension at a declination of -55:56:30. Right: a slice through declination at a right ascension of 06:58:21.2. These slices passes through the X-ray centroid of the smaller cluster (the bullet). In the slice through right ascension an increase in the radio halo brightness is visible close to the position of the bullet (the bullet is clearly seen in the X-ray profile at a declination of 104.58$^\circ$). The slice through right ascension passes also passes close ($6\arcsec$) to the peak in the centroid of the X-ray emission from the main cluster at 104.62$^\circ$ and through the peak the radio halo emission at 104.63$^\circ$.

    \includegraphics[width=5.8cm]{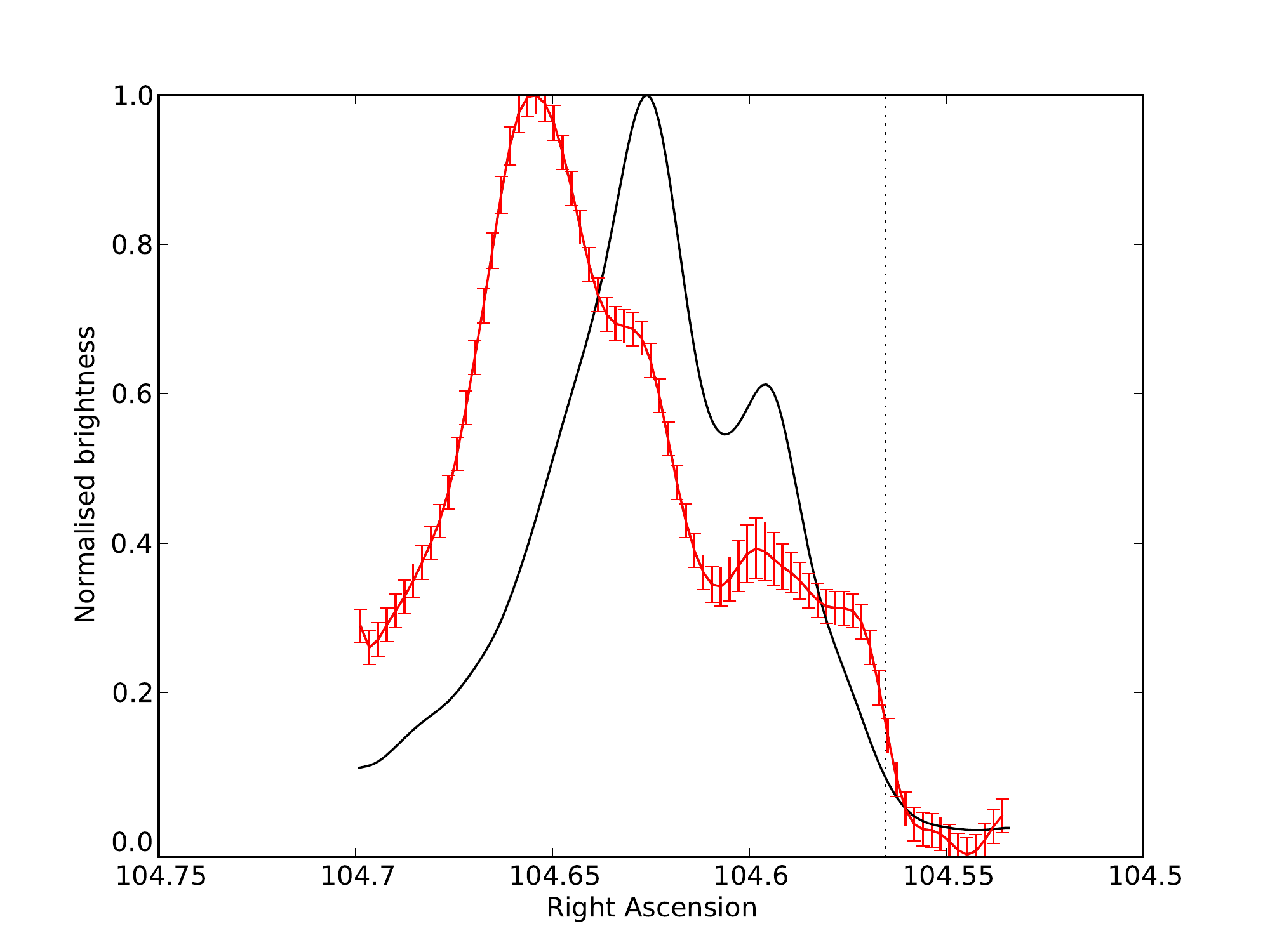} \includegraphics[width=5.8cm]{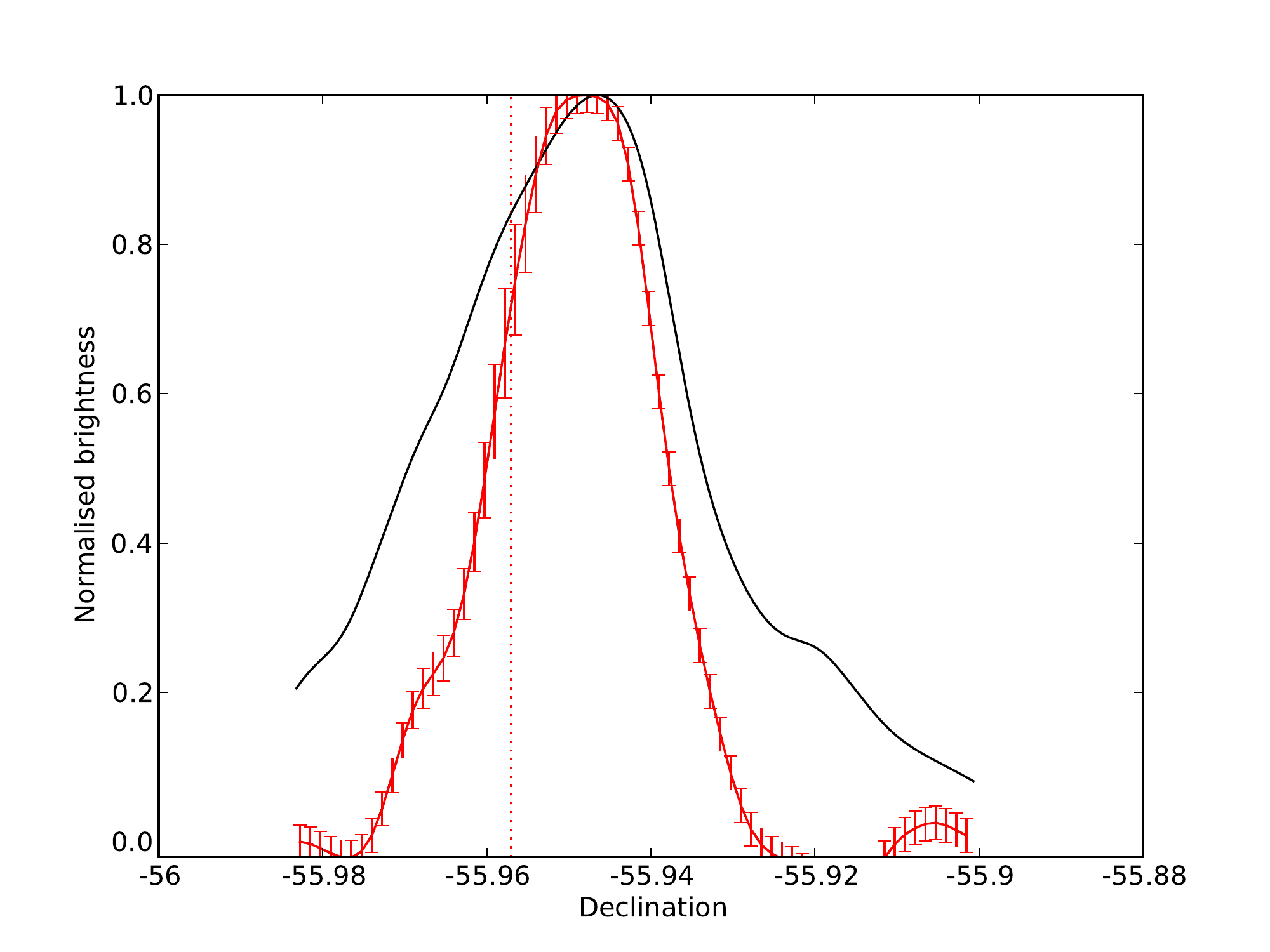} 
   
    Left: a slice through right ascension at a declination of -55:57:00. Right: a slice through declination at a right ascension of 06:58:36. 
    
   \caption{One-dimensional brightness profiles of the X-ray and radio emission. The X-ray emission is shown by solid black lines, the radio emission is shown with solid red lines and the error bars are $\pm (30 + \rm{R}  \exp^{\frac{-X^2}{2\sigma^2}})\mu$Jy/beam, where R is the residual flux-density after source subtraction (see Table \ref{tab:ATCA_SOURCES}), $X$ is the separation of the source from the position in the slice and $\sigma$ is the synthesised beam standard deviation (FWHM=23.3$\arcsec$). The errors shown are thus not simple random errors but also show the maximum contamination from the residuals that remain after \textsc{clean} component subtraction which are not random but will increase the flux. The locations of sources that contribute more than 30$\mu$Jy/beam in errors are shown with red dotted vertical lines. For the slices at constant declination the location of the X-ray detected shock front is shown by a grey vertical dotted line. The X-ray emission has been convolved with a 20$\arcsec$ Gaussian to aid comparison with the low resolution radio image. Both the X-ray and radio brightness profiles have been divided by a normalisation factor so that they peak at 1.0. On the left from top to bottom the radio normalisation factors are 0.6, 1.5 and 1.3mJy/beam and the corresponding X-ray normalisation factors are 1.7, 3.0, 2.4$\times10^{-4}$cts s$^{-1}$pixel$^{-1}$. On the right from top to bottom the radio normalisation factors are 1.3, 0.7, 1.3mJy/beam and the corresponding X-ray normalisation factors are 2.8, 3.0 and 1.2$\times10^{-4}$cts s$^{-1}$pixel$^{-1}$.}
   \label{1d-slices}
\end{figure*}

Along the east-west axis, we have taken three one-dimensional slices through the X-ray and radio emission (see the left hand side of Figure \ref{1d-slices}), and these show that along this axis the extent and brightness profile of the X-ray emission is similar to that seen in the radio. However, in the north-south direction the extent  of the radio halo ($\approx$130$\arcsec$) is significantly less than that of the X-ray emission (see the top and bottom panels on the right hand side of Figure \ref{1d-slices}). In this cluster system, the smaller cluster (the bullet) has passed through a larger cluster 0.1-0.2\,Gyrs ago (\citealt{Markevitch_2006}); if merger turbulence were not yet widespread throughout the intra-cluster medium, this could explain the lack of radio emission in the northern and southern extremities of the cluster. In the direction of the merger, where we would expect significant turbulence, we observe that the halo appears to trace out the path taken by the bullet and extends to the eastern and western extremities of the cluster. The morphology of the halo therefore appears to be consistent with what would be expected from a radio halo still in formation and formed as a consequence of a recent merging event. Furthermore, the observed extension is reproduced in a recent suite of simulations by \cite{Donnert_2013}, who simulated a radio halo in a system similar to the bullet cluster and found that turbulence behind the infalling cluster accelerates a population of seed non-thermal electrons to produce a stream of relativistic synchrotron emitting electrons.

We find that the peak of the radio halo emission (06:58:31.6 -55:56:29.5) is coincident with the X-ray centroid of the main cluster (see the centre left panel of Figure \ref{1d-slices}) but 35$\arcsec$ (150\,kpc) north-west of the weak-lensing centroid (where we have used the X-ray and weak-lensing centroids given by \cite{Clowe_2006} as 06:58:30.2 -55:56:35.9 and 06:58:35.3 -55:56:56.3). However, the bright X-ray structure of the main cluster component is primarily extended in the north-south direction, but the bright region of the radio halo extends primarily to the south-west towards the hottest region in the cluster (excluding the temperatures derived at the shock) and the weak-lensing mass reconstruction centroid. Figure \ref{1d-slices} demonstrates that this misalignment between the X-ray and radio halo emission is predominantly in the direction of the cluster merger; in the direction perpendicular to this the X-ray and radio brightness emission is better aligned. Simulations presented by \cite{Donnert_2013}, where the origin of radio halo emission is turbulent-acceleration, also show a misalignment between the radio and X-ray emission in the direction of the cluster merger.

Constraining our search of a radio halo peak to the region surrounding the bullet, we find a radio halo peak at 06:58:23.6 -55:56:43.0 which is offset by 20$\arcsec$ (88\,kpc) eastwards from the centroid of the X-ray emission from the bullet (06:58:21.2 -55:56:30.0) and 65$\arcsec$ (270\,kpc) eastwards from the centroid of the weak-lensing mass reconstruction of the bullet (06:58:16.0 -55:56:35.1). The small radio brightness peak can also be seen in the centre left panel of Figure \ref{1d-slices}, although there are two subtracted sources whose residuals introduce additional uncertainty, the faint peak appears real but is a low significance detection. We note that the errors shown in Figure \ref{1d-slices} are not simply random errors; instead they consist of 30$\mu$Jy/beam random errors plus a maximum contamination from the residuals that remain after \textsc{clean} component subtraction (in this region these are around $50\mu$Jy/beam), which will increase the flux in this region. The  presence of a second peak in the radio halo brightness suggests that the halo could have two components, one associated with the main cluster and another associated with the bullet.

On the western tip of the radio halo the north-south extent is small but initially it increases smoothly as a function of distance from the western tip. The north-south extent reaches a maximum of $\approx$ 230$\arcsec$ about $\approx 1 \arcmin$ from the western tip after which it rapidly decreases to about $\approx$ 130$\arcsec$ and maintains this north-south extent across the rest of the halo. This maximum in the north-south extension of the halo is approximately equal in extent to the X-ray emission, and in this region the brightness profiles of the X-ray and radio emission are similar (see Figure \ref{fig-bullet-cluster} and the centre right panel of Figure \ref{1d-slices}). In contrast, as previously mentioned, the rest of the cluster has an X-ray morphology that extends beyond the radio halo emission in the north-south direction. These northern and southern extensions of the radio halo are faint, and are only visible close to the shock that precedes the bullet as it propagates through the larger cluster. A possibility is that the shock is partially responsible for the emission and indeed, the distinctive shape of the western edge of the halo mimics that of the X-ray detected shock front with Mach number $M=3.0\pm0.4$ (\citealt{Markevitch_2006}). To demonstrate this morphological similarity, in Figure \ref{fig-bullet-cluster} we have overlaid the X-ray shock on the radio halo emission and see excellent correspondence. The efficiency with which shocks at this low Mach number can accelerate electrons or protons (which could then decay to produce the relativistic electrons) is unknown (e.g. \citealt{Kang_2007}, \citealt{Riquelme_2011} and \citealt{Kang_2012}). However, our observations of the radio halo of 1E 0657-55.8 indicate that merger shocks play a major role in the formation of radio halos and show that even this low Mach number shock can instigate synchrotron emission.

\subsection{Correlations}

\begin{figure}
 \includegraphics[height=6.0cm]{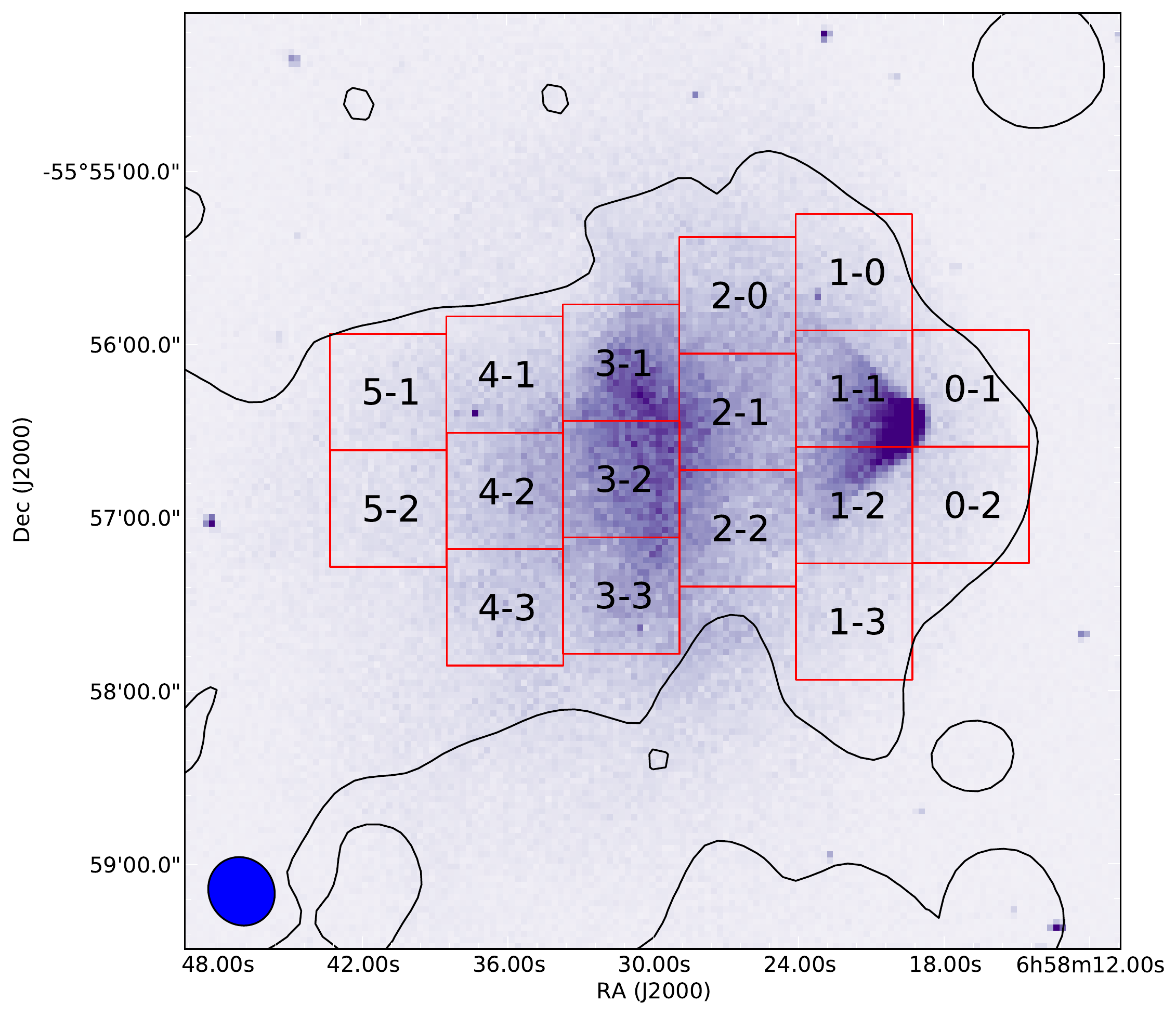} 
   \caption{The labelled 40$\arcsec \times 40\arcsec$ red boxes show the regions used for a point-to-point comparison between images. The X-ray emission is shown in colour and the solid line is the $25\mu$Jy/beam contour level of the low resolution radio image (synthesised beam FWHM equals 23.3$\arcsec$).}
    \label{grid-dray-radio}
\end{figure}

\begin{figure*}
\centering
     \includegraphics[width=7.8cm]{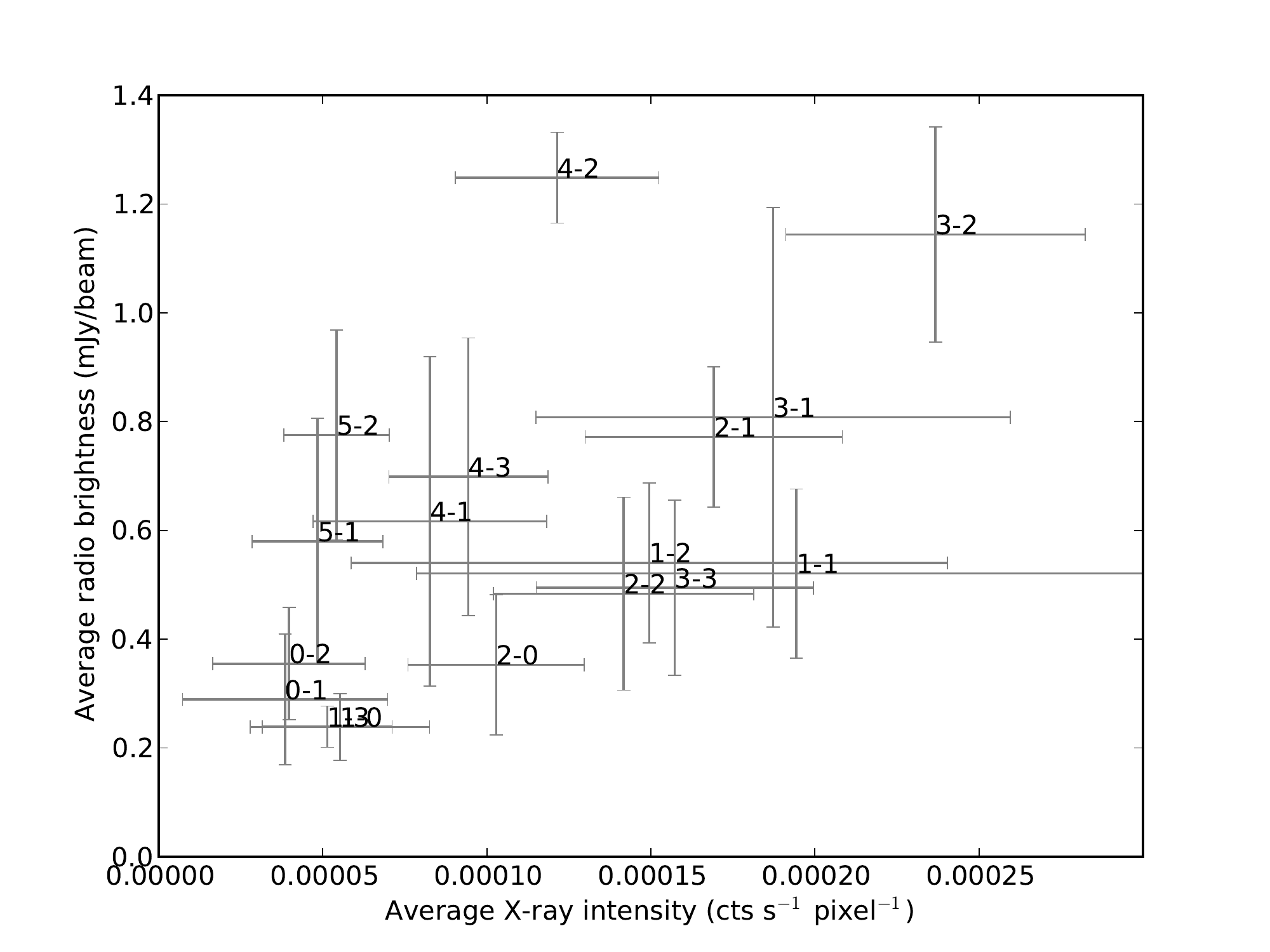} \includegraphics[width=7.8cm]{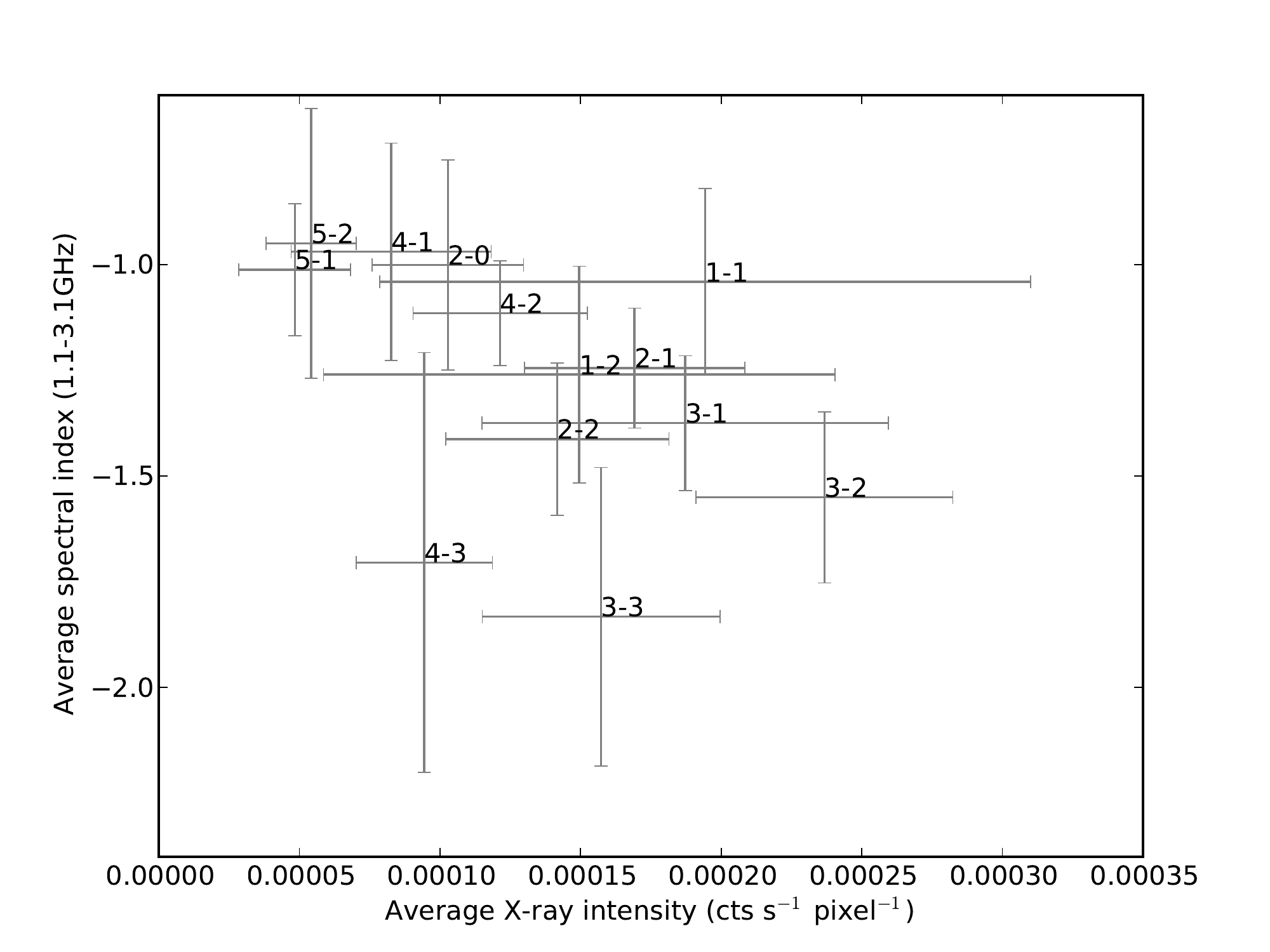} 
   \includegraphics[width=7.8cm]{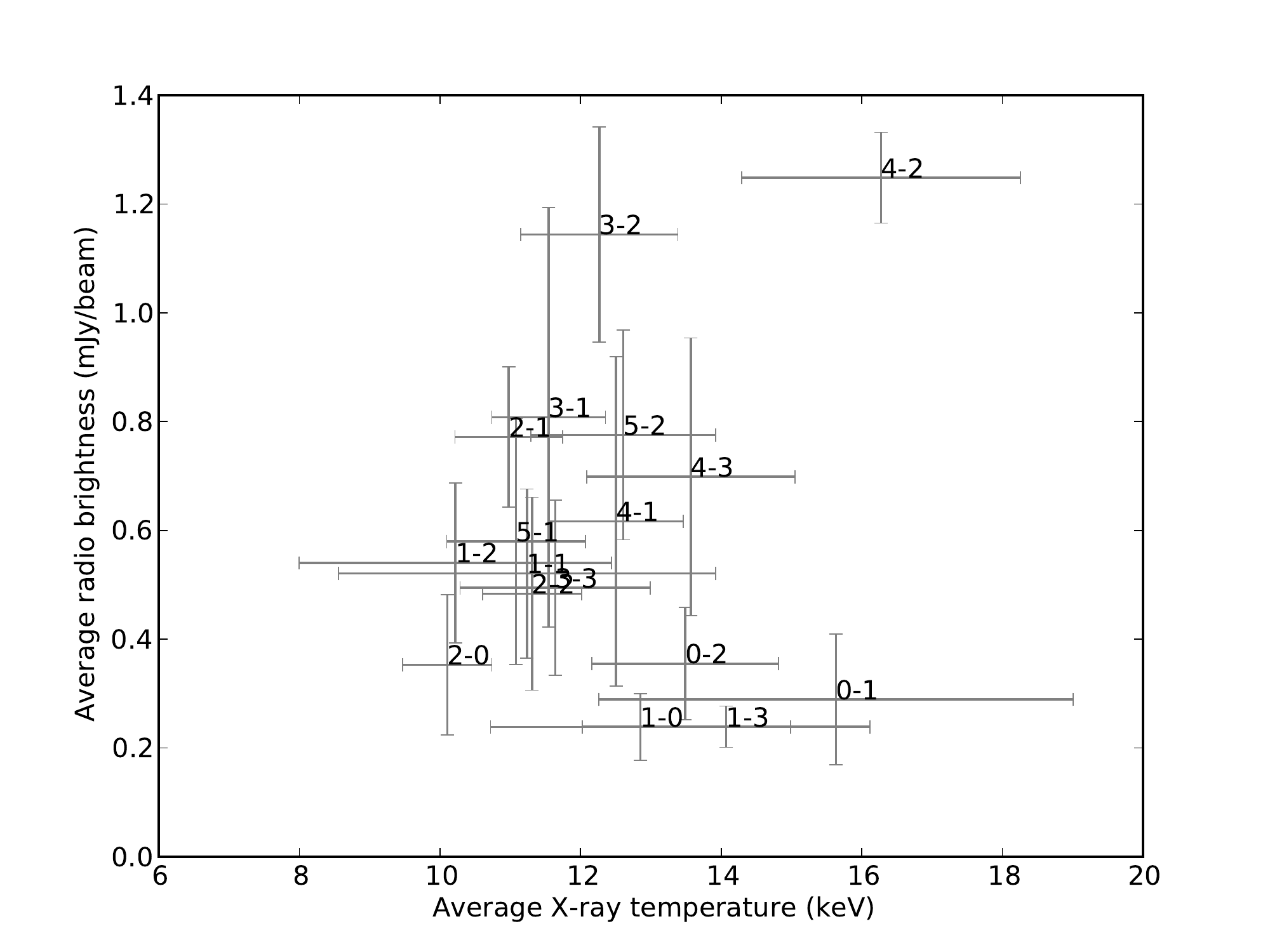} \label{decslice0}  \includegraphics[width=7.8cm]{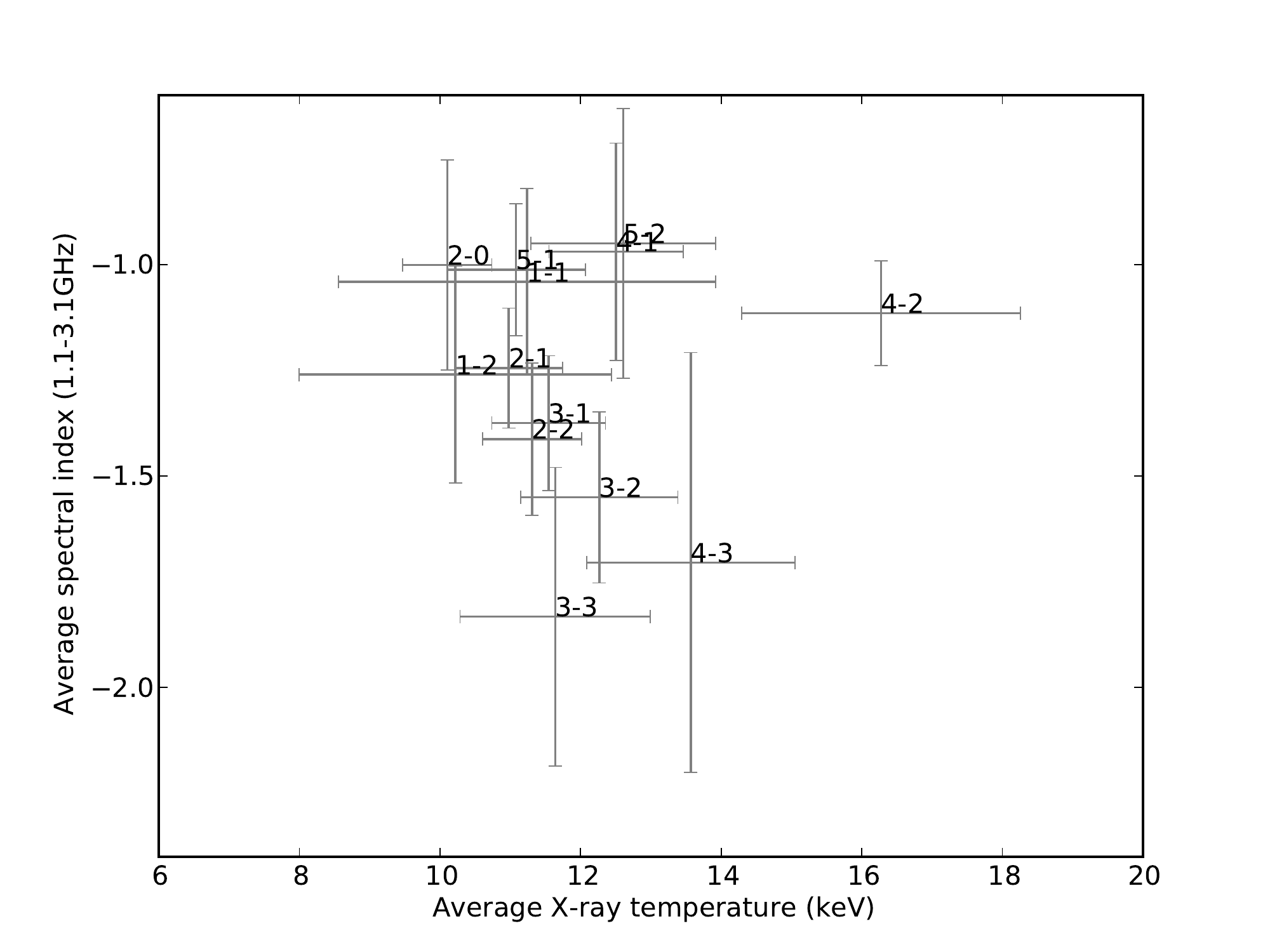} \label{decslice0}  
   \includegraphics[width=7.8cm]{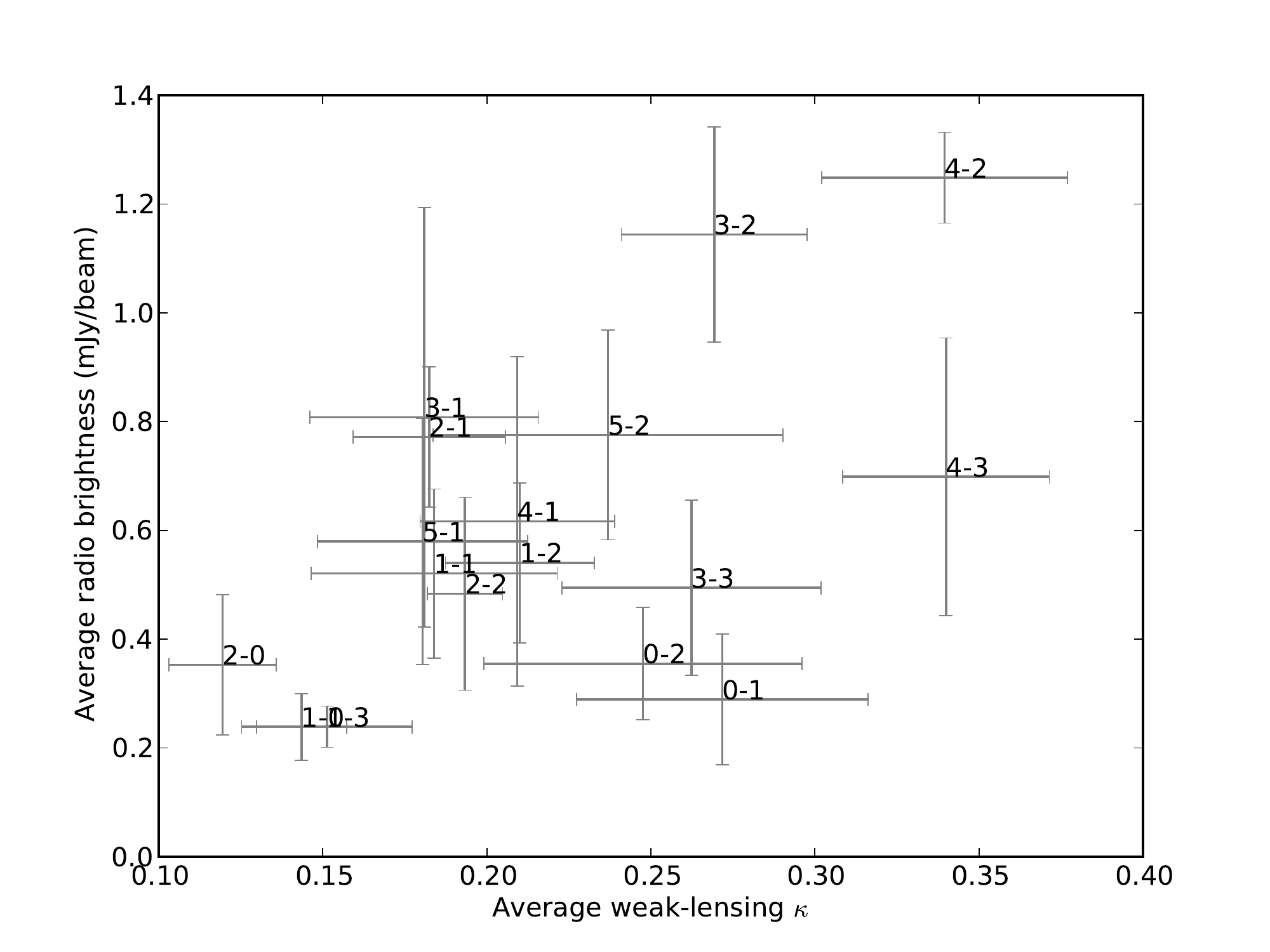} \label{decslice0}\includegraphics[width=7.8cm]{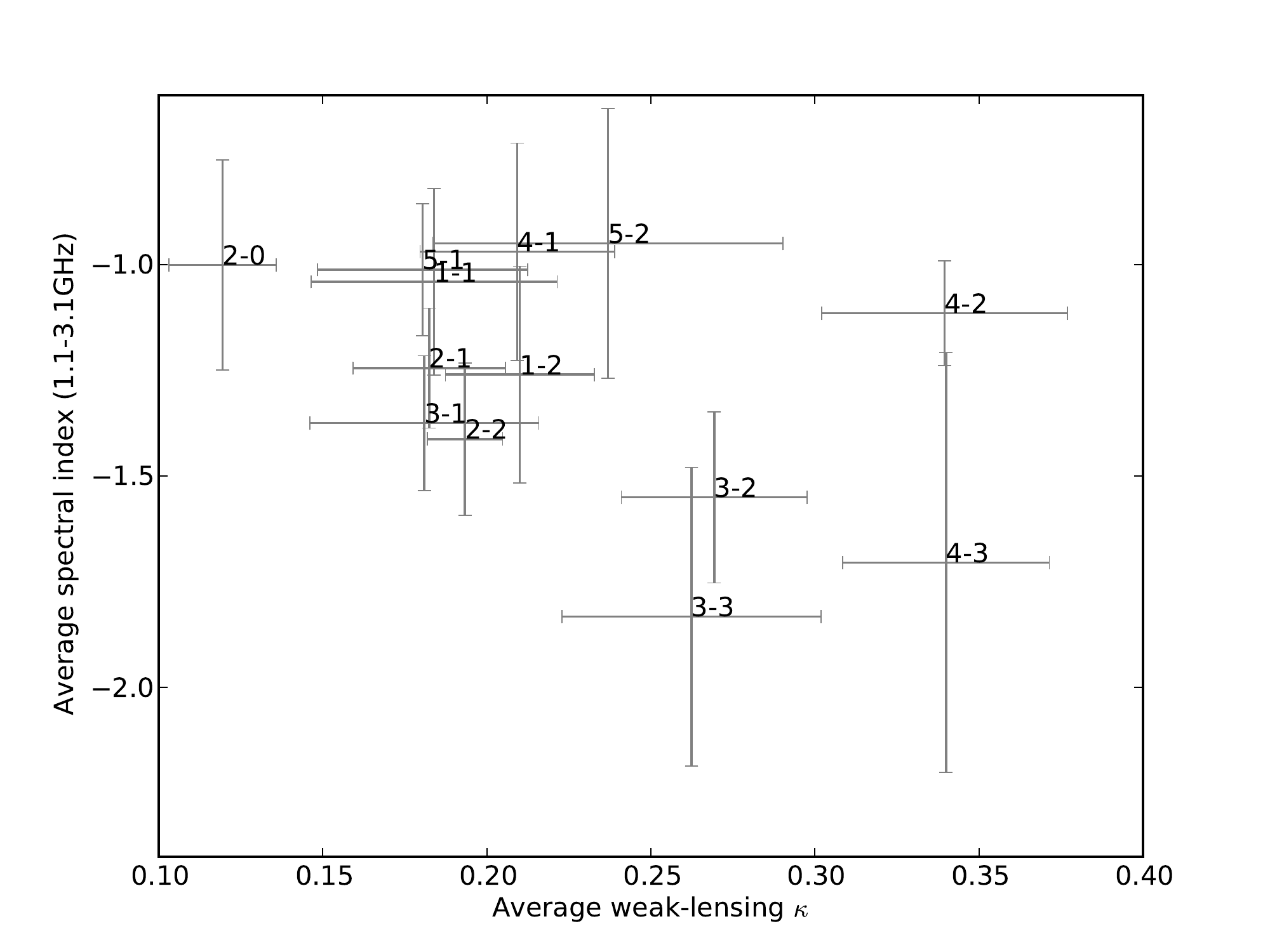} 
   \caption{Point-to-point comparison of the radio image and wide-band spectral index map with the X-ray and weak lensing images. Each point is labelled corresponding to its position in the grid shown in  Figure \ref{grid-dray-radio}.}
   \label{fig:point-to-point}
\end{figure*}

It has been observed that there is a trend for clusters with high X-ray temperatures to have flatter radio halo spectra (see \citealt{Feretti_2004}, \citealt{Giovannini_2009} and \citealt{Venturi_2013}). Furthermore, several studies (e.g. \citealt{Feretti_2001}, \citealt{Govoni_2001} and \citealt{Giacintucci_2005}) have found that for regions within a single cluster, there is a positive correlation between radio halo luminosity and X-ray luminosity. We have performed a point-to-point comparison of our radio brightness and spectral index images with the X-ray and weak-lensing mass reconstruction images, using the grid shown in Figure \ref{grid-dray-radio}. Using the same procedure as \cite{Govoni_2001}, we estimate the errors from the standard deviation of pixels within the grid cells. Figure \ref{fig:point-to-point} shows the observed correspondence between the X-ray, weak-lensing and radio properties. For each of these plots, we calculate the Spearman's rank correlation coefficient ($\rho$), the non-parametric nature of this correlation coefficient allows us to identify whether there is a monotonic relationship between variables. From the correlation coefficient we determine the z-score ($z_{s}$), which can be interpreted as the Gaussian significance of correlation:
\begin{equation}
z_{s} = F(\rho) \sqrt{\frac{N-3}{1.06}},
\label{eqn:z-score}
\end{equation}
where $F(\rho)$ is the Fisher transformation of the Spearman rank correlation coefficient and N is the number of data pairs. To propagate the uncertainty from our measurement values, we use a Monte-Carlo analysis where we resample each measurement from a Gaussian distribution centred on the mean with a standard deviation equal to the measurement error. These resampled values form a new measurement set for which we calculate $\rho$ and  $z_{s}$. Repeating the procedure a statistically robust number of times produces a Gaussian distribution of z-scores from which we calculate the mean, which is our significance of correlation, and the standard deviation, which is our error on this significance of correlation. The results are given in Table \ref{tab:z-scores}, and demonstrate that we detect no significant correlations between the radio and weak-lensing or radio and X-ray parameters.  
Given that some of these correlations have been observed in some clusters, in particular the relation between X-ray and radio halo brightness which can be a tight correlation (see e.g. \citealt{Govoni_2001}), the lack of a detected correlation indicates that this radio halo may be in a different state from those obeying the scaling relation which typically have a more circular appearance. The absence of the scaling relation could be due to a number of things, such as: the extreme merging state of this cluster leading to a complex distribution of thermal and non-thermal gas; the large variations in temperature across the cluster; projection effects; or that the radio halo may still be forming and is in a different stage of its evolution than those that obey the scaling relations. Alternatively, there could be two components to this halo, each obeying a scaling relation; and the observed signal showing the combined emission not strictly following the relations. However, this second hypothesis seems unlikely, because if we remove the data from the three western grid columns in Figure \ref{grid-dray-radio} from the Spearman rank calculations, thus leaving just the regions close to the main cluster, we find no improvement in the significance of correlation.

\begin{table}
\caption{Derived z-scores and Spearman rank correlation coefficients together with their corresponding errors for the plots presented in Figure \ref{fig:point-to-point}. In brackets we present the same parameters but derived from the narrow-band spectral index image (Figure \ref{traditional-spec}) rather than the wide-band spectral index image (Figure \ref{fig-bullet-cluster}). No significant correlation is observed between the radio and weak-lensing or radio and X-ray parameters.}
 \centering
 \label{tab:z-scores}
\begin{tabular}{lccccc}
\hline 
x-axis & y-axis & $z_s$ & $\rho$ \\  \hline
X-ray intensity & Radio brightness & $1.6\pm0.7$ &$0.4\pm0.2$\\
X-ray intensity & Spectral index & $-1.2\pm0.8$ &$-0.4\pm0.2$\\\
                      &                       & ($-0.8\pm0.8$) &($-0.3\pm0.2$)\\\
X-ray temperature & Radio brigtness & $0.0\pm0.7$ &$0.0\pm0.2$\\\
X-ray temperature & Spectral index & $-0.1\pm0.8$ &$0.0\pm0.3$\\\
                      &                       & ($0.1\pm0.7$) &$(0.0\pm0.3$)\\\
Weak lensing & Radio brightness & $1.5\pm0.7$ &$0.3\pm0.2$\ \\
Weak lensing & Spectral index & $-1.0\pm0.8$ &$-0.3\pm0.2$\ \\
                      &                       & ($-0.3\pm0.8$) &($-0.1\pm0.3$)\ \\ \hline
 \end{tabular}
\end{table}

\subsection{Polarisation}

To date, polarisation has been very hard to detect in radio halos, with the only detected polarised emission being filamentary structures in the clusters A2255 and MACS J0717+3745 (\citealt{Govoni_2005}, \citealt{Pizzo_2011} and \citealt{Bonafede_2009}). Cluster-wide polarised emission that is indisputably associated with a radio halo has not been detected in any cluster. Beam depolarisation is thought to be a major cause for these non-detections. This agrees with recent simulations by \cite{Govoni_2013} (Figure 4 of that paper), who show that the fractional polarisation at the centre of a simulated cluster is 2\%, 3\% and 7\% when observed with resolutions of 50$\arcsec$, 30$\arcsec$ and 10$\arcsec$, respectively. The simulations predict that the centre of the cluster has the lowest fractional polarisation (due to internal depolarisation), but as it is the brightest region of the halo in Stokes I, it is still the region of highest polarised surface brightness. In the case of the bullet cluster, we may also expect the shock on the western edge to align the magnetic field and for the observable fractional polarisation in this region to be higher than usual (radio relics, which are thought to be caused by shocks, are often 20-30\% polarised). However, we have not been able to detect any polarised emission from within the region of the radio halo. Our 5$\sigma_{Q,U}$ limits on the fractional polarisation of 13\% when observed with 21$\arcsec$ resolution and 16\% at $17\arcsec$ resolution are still significantly higher than  \cite{Govoni_2013} predictions of the radio halo polarised emission. These limits are from wide-band 1.1-3.1\,GHz images, whereas the \cite{Liang_2000} limits (3$\sigma$ limits at 20\%, 6.5\% and 1.4\% at resolutions of 10$\arcsec$, 20$\arcsec$ and 60$\arcsec$, respectively) are at 1.4\,GHz. Our 1.4\,GHz limits are comparable to those of \cite{Liang_2000}.

\section{Conclusions}

We have analysed deep, wide-band, polarimetric, radio observations of the bullet cluster. After removing the contamination from radio sources, we have presented images of the bullet cluster radio halo that have significantly better sensitivity than previously published images of this halo. From these observations we have been able to characterise the morphology and spectral properties of the radio halo, but have been unable to detect polarised emission from any regions of the halo. From our study we have found the following:
\begin{itemize}
\item We determined an integrated 1.3\,GHz flux-density of 52.5$\pm$2.1\,mJy from the region of significant radio halo emission. In this region our measurements are in good agreement with the \cite{Liang_2000} measurements from a comparable region. However, when integrating over a region much larger than the area of significant emission we measure an integrated 1.3\,GHz flux density of 56.4$\pm$2.3\,mJy which is lower than the large area \cite{Liang_2000} measurement of 78$\pm$5\,mJy.
\item Our measured 1.3\,GHz to 3.1\,GHz spectral index ($-1.57\pm0.05$) is steeper than the 0.8\,GHz to 8.8\,GHz spectral index measured by \cite{Liang_2000} (-1.3$\pm$0.1). 
\item The radio emission extends as far as the observed X-ray emission in the direction of the cluster merger, but perpendicular to this axis, the extent of the radio emission is smaller than that of the X-ray emission.
\item We find evidence for two peaks in the radio halo emission. One is coincident with the X-ray centroid of the main cluster and another is close to the bullet. 
\item The centroids of surface brightness slices through the halo and X-ray structures are better aligned in the direction perpendicular to the merging axis than in the direction of the cluster merger.
\item There is a distinctive edge to the bullet cluster radio halo, which is coincident with the bow shock detected in deep X-ray observations.
\item The radio intensity and spectral index distribution do not have tight correspondences with the X-ray brightness, X-ray temperature or weak-lensing mass reconstruction.
\item The radio spectral index varies across the cluster but there is no clear trend.
\item We do not detect any polarised emission from the radio halo. Our 5$\sigma_{Q,U}$ upper limits on the fractional polarisation in the centre of the cluster over the 1.1-3.1\,GHz band are 13\% at a resolution of 21$\arcsec$. 
\end{itemize}

\section{Acknowledgements}

We thank the anonymous referee for their comments. The Australia Telescope Compact Array is part of the Australia Telescope National Facility which is funded by the Commonwealth of Australia for operation as a National Facility managed by CSIRO. We thank Maxim Markevitch for useful discussions and for kindly providing X-ray brightness and temperature images. We thank Amy Kimball for comments which improved this paper. B.M.G. acknowledges the support of Australian Laureate Fellowship FL100100114 from the Australian Research Council.

\bsp
\label{lastpage}

\end{document}